\documentclass[preprints,article,accept,moreauthors,pdftex]{Definitions/mdpi}

	\DeclareGraphicsExtensions{.pdf,.jpeg,.png}
	\usepackage{array}
	\usepackage{mdwmath}
	\usepackage{mdwtab}
	\usepackage{eqparbox}
	\usepackage{url}
	\usepackage{tabu,multirow}
	\usepackage{soul,color}
	\usepackage{subcaption}
	\usepackage{euflag}
    \usepackage{stfloats}
    \usepackage{longtable}
    \newcolumntype{C}[1]{>{\centering\arraybackslash}m{#1}} 

\firstpage{1} 
\pubvolume{xx}
\issuenum{1}
\articlenumber{5}
\pubyear{2019}
\copyrightyear{2019}
\history{Received: date; Accepted: date; Published: date}

\Title{A Review of Cyber-Ranges and Test-Beds: Current and Future Trends}

\Author{Elochukwu Ukwandu $^{7}$\orcidA{}, 
Mohamed Amine Ben Farah $^{1}$\orcidB{}, 
Hanan Hindy $^{4}$\orcidC{}, 
David Brosset $^{2}$\orcidD{}, 
Dimitris Kavallieros $^{5,6}$\orcidE{}, 
Robert Atkinson $^{1}$\orcidF{}, 
Christos Tachtatzis $^{1}$\orcidG{}, 
Miroslav Bures $^{3}$\orcidH{}, 
Ivan Andonovic $^{1}$\orcidI{}, 
and Xavier Bellekens $^{1}$\orcidJ{} }

\AuthorNames{Elochukwu Ukwandu, Mohamed Amine Ben Farah, Hanan Hindy, David Brosset, Dimitris Kavallieros, Robert Atkinson, Christos Tachtatzis, Miroslav Bures, Ivan Andonovic and Xavier Bellekens}

\address{%
$^{1}$ \quad Dept. of Electronic and Electrical Engineering, University of Strathclyde, Glasgow, United Kingdom; {mohamed.ben-farah,robert.atkinson,christos.tachtatzis,ivan.andonovic,xavier.bellekens}@strath.ac.uk\\
$^{2}$ \quad  Naval Academy Research Institute, Arts et M\'{e}tiers Institute of Technology, France; david.brosset@ecole-navale.fr \\
$^{3}$ \quad  Department of Computer Science, FEE, Czech Technical University in Prague; buresm3@fel.cvut.cz \\
$^{4}$ \quad Department of Cyber-Security, Abertay University, United Kingdom ; 1704847@abertay.ac.uk \\
$^{5}$ \quad The Center for Security Studies (KEMEA);\\
$^{6}$ \quad University of Peloponnese, Department of Informatics and Telecommunications; \\
$^{7}$ \quad Dept. of Computer Science, Cardiff School of Technologies, Cardiff Metropolitan University, Cardiff, Wales, United Kingdom; eaukwandu@cardiffmet.ac.uk \\ 
}

\corres{Correspondence: xavier.bellekens@strath.ac.uk}

\abstract{Cyber situational awareness has been proven to be of value in forming a comprehensive understanding of threats and vulnerabilities within organisations, as the degree of exposure is governed by the prevailing levels of cyber-hygiene and established processes. A more accurate assessment of the security provision informs on the most vulnerable environments that necessitate more diligent management. The rapid proliferation in the automation of cyber-attacks is reducing the gap between information and operational technologies and the need to review the current levels of robustness against new sophisticated cyber-attacks, trends, technologies and mitigation countermeasures has become pressing. A deeper characterisation is also the basis with which to predict future vulnerabilities in turn guiding the most appropriate deployment technologies. Thus, refreshing established practices and the scope of the training to support the decision making of users and operators. The foundation of the training provision is the use of Cyber-Ranges (CRs) and Test-Beds (TBs), platforms/tools that help inculcate a deeper understanding of the evolution of an attack and the methodology to deploy the most impactful countermeasures to arrest breaches. In this paper, an evaluation of documented CR and TB platforms is evaluated. CRs and TBs are segmented by type, technology, threat scenarios, applications and the scope of attainable training. To enrich the analysis of documented CR and TB research and cap the study, a taxonomy is developed to provide a broader comprehension of the future of CRs and TBs. The taxonomy elaborates on the CRs/TBs different dimensions, as well as, highlighting a diminishing differentiation between application areas. }


\begin{document}

\maketitle

\section{\textbf{Introduction}}
\label{Section: Introduction}
In the recent past, a proliferation in the number and complexity of cyber-security incidents with deeper consequences is evident as attackers become more skilled, sophisticated and persistent. The extent of cyber-incidents targeting critical infrastructures and the public has been fueled further by global events such as the recent COVID-19 pandemic, impacting a plethora of organisations and fueling a clear and immediate need for increased cyber-situational awareness~\cite{lallie2020cyber}. Furthermore, in the recent past, the cyber-security industry has undergone a significant shift in respect of acknowledging the importance of security training of users, transitioning from ``users are the weak link of cyber-security" towards ``users can be trained like muscles hence, improving a company's overall security posture". The evolving change of stance is a fundamental trigger for change in cyber-security procedures, in turn stimulating a growing demand for training platforms. Cyber-Ranges~(CR) and Test-Beds~(TB) are the foundation for the creation and emulation of adaptable Information Technology~(IT) and Operational Technology~(OT) networks, respectively. Scenarios replicating a spectrum of cyber-attacks can be established, enhancing the training of operators and users within recognisable environments in the identification of, and mitigation strategies to arrest cyber-breaches. Training in an emulated environment accelerates effective learning of best practice and the `real-time' dynamic interaction promotes a deeper understanding of the consequences of any action. CR/TB facilitate the establishment of an extended range of attack scenarios with varying levels of complexity, governed by the stage of training. Groups of users can also train on a remotely accessible platform to define, optimise and evaluate the impact of a coordinated response to cyber-attacks, e.g. `blue team', `red team', back and `front office' of (say) a bank. Group training involving multiple teams comprising varying knowledge sets enhance the cyber-situational awareness of the organisation and improve the response time to identify and arrest a cyber-attack.

In response to the pressuring demand to respond to exponentially evolving cyber-attacks, this paper presents an  extensive and thorough analysis of CRs and TBs based on the recent prominent research and manuscripts. To the best of the authors knowledge, this thorough analysis is not available in the literature, thus limiting the presence of an adequate resource for researchers. Moreover, to complete the study, two taxonomies targeted towards the different dimensions of CRs and TBs are developed and presented in this manuscript.  

The remainder of the paper is organised as follows. Section~\ref{Section: Methodology} details the methodology applied to execute on a review of the state-of-the-art in CRs/TBs; Section~\ref{Section: RelatedWorks} presents a summary of existing knowledge in the disciplines. Section~\ref{Section: SystematicReview} provides a critical assessment on reported CRs/TBs classified by the domain of applications, user classes, method of experimentation and implementation. Section~\ref{Section: ScenariosandApplications} covers the scenarios and applications of CRs/TBs. Section~\ref{Section: AnalysisandTaxonomies} focuses on CR/TB taxonomies informed by the findings of the review. Section~\ref{Section: TrainingMethods} describes the training methods implemented through CRs/TBs with the dynamics and methods used in analysing threats summarised in Section~\ref{Section: ThreatDynamicsandAnalyses}. Section~\ref{Section: TheFutureofCRsandTBs} elaborates on the future evolution and use of CRs/TBs, providing evidence of the narrowing gap between their different application areas. Conclusions are drawn in Section~\ref{Section: Conclusions}.

\section{\textbf{Methodology}}
This section provides detailed of how the review was conducted and method used.

\label{Section: Methodology}
\subsection{\textbf{Overview}}
The review presented here adopts a high-level systematic methodology based on planning, selection, extraction, and execution in line with the guidance prescribed by Okoli and Schabram in~\cite{okoli2010guide} and Okoli in~\cite{okoli2015guide}. The review is stand-alone focusing on existing knowledge, evaluation, and synthesis in the domains of CR and TB, the principle aim being to provide evidence of the growing density of cyber-attacks events harnessing the features of Artificial Intelligence~(AI) and Bio-inspired systems to automate attack processes with increasing levels of stealth and sophistication in both landscape and execution. Furthermore, the increasing degrees of network inter-connectivity as a consequence, for example, of emerging Industry 4.0 `smart-everything' scenarios and the concomitant changes in the dynamics and scope of the threat surface, translates into a major challenge in determining evolving cyber situational awareness for researchers, educators, and trainers. The prediction of future trends, scenarios, and possible application areas using current operational environments presents significant challenges. Therefore, this paper - a study with these aims has not been reported to date - projects the training requirements for cyber situational awareness within these evolving infrastructures utilising the existing knowledge within the literature and current sector practices as the seed. 

\subsection{\textbf{Aim and Objectives}}
The literature review aims to identify and analyse the current state-of-the-art in the use and applications of CRs/TBs within cyber-security training and map the range of applications provisioned by these platforms. The objectives are as follows:
\begin{itemize}
    \item Classifying CRs and TBs.
    \item To identify and review state-of-the-art trends, scenarios, applications, and training methods.
    \item Identifying threat dynamics and analysis methods.
    \item To leverage the knowledge reported to date as the seed to provide insights into emerging future trends, scenarios, technologies, application areas, and training methods fit aligned with the evolution of data-driven practices such as Industry 4.0
    \item To help equip cyber-security professionals, educators, and trainers with the relevant skills to combat cyber-threats in next generation highly inter-connected, multi-domain infrastructures
    \item Aims to establish new taxonomies for future CRs/TBs informed by the findings from the survey.\\
\end{itemize}

\subsection{\textbf{CRs and TBs Survey}}
A comprehensive literature review of the state-of-the-art in CRs/TBs disciplines was carried out to establish a reference of current platform features and training tools, the foundation for the development of the main contributions presented in the paper.

\subsubsection{\textbf{Classification and Research Criteria}}
As CR and TB migrate towards convergence, the literature search used the following keywords to surface the most relevant publications: \\
1. ``Cyber-ranges" +(``Military" + ``Defense" + ``Intelligence") or (``Industry" + ``Commercial") or (``Education" + ``Research")\\
2. ``Test-bed"+ (``IoT" or ``Smart Grid" or ``Cloud") + cyber\\
Furthermore, the review only considered papers published within the last 5 years of 2015-2020.

\subsubsection{\textbf{Selection Criteria}}
Searches in five databases were executed: ScienceDirect, IEEE Explore, Springer,  Wiley, and ACM. However,  fundamental research relevant to the study outwith the specified search period were taken into consideration. The graphs presented in Figures~\ref{fig:1},~\ref{fig:2} and~\ref{fig:3} depict the evolution of the number of publications from 2015 to 2020 on Test-beds in the Internet-of-Things (IoT), Smart Grid and Cloud disciplines. Increases in the number of publications for all domains is clearly evident, demonstrating extensive research in cyber-security and Test-beds~\cite{10.1007/978-3-030-58768-0_6}.

\subsubsection{\textbf{Extraction Criteria and Results}}
The review was restricted further to consider only the current dominant area of application for TBs - Smart Grids - and the future area of IoT/smart devices, driven by the goal of predicting future requirements seeded by the current state-of-the-art.

\begin{figure}[tb!]
\centering
    \begin{subfigure}{0.32\textwidth}
    \centering
    \includegraphics[width=\linewidth]{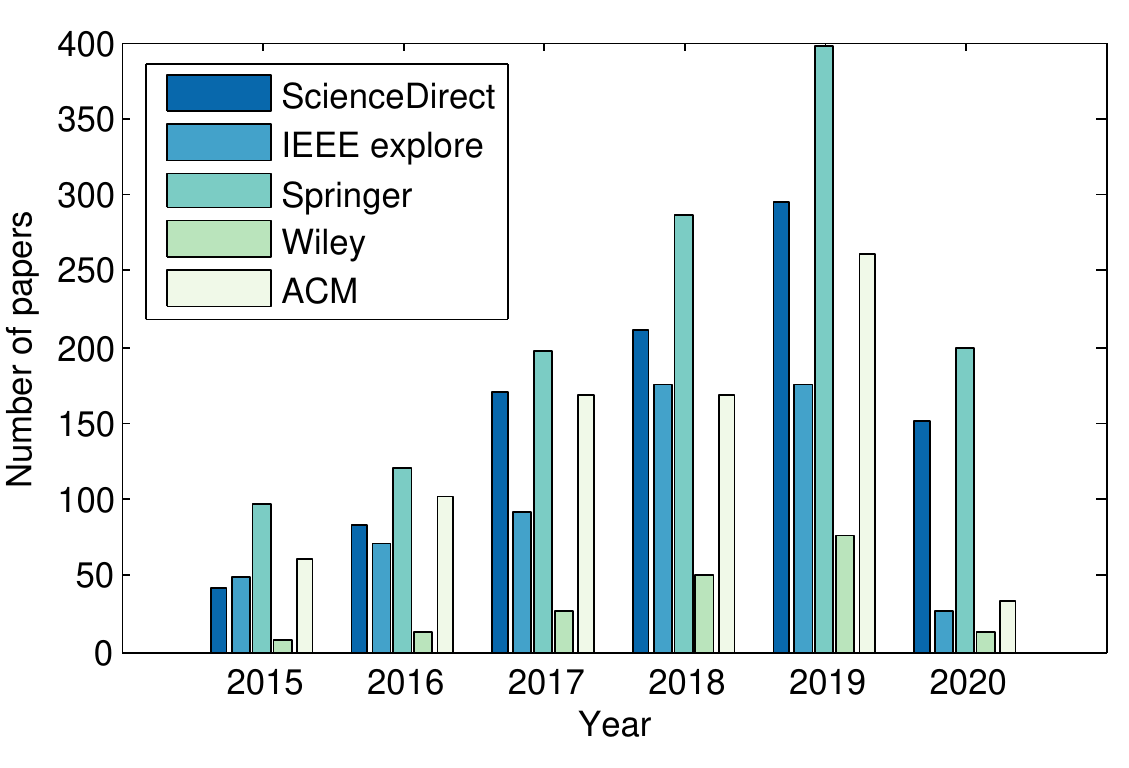}
    \caption{TBs in IoT-based papers}
    \label{fig:1}
  \end{subfigure}
  \begin{subfigure}{0.32\textwidth}
    \centering
    \includegraphics[width=\linewidth]{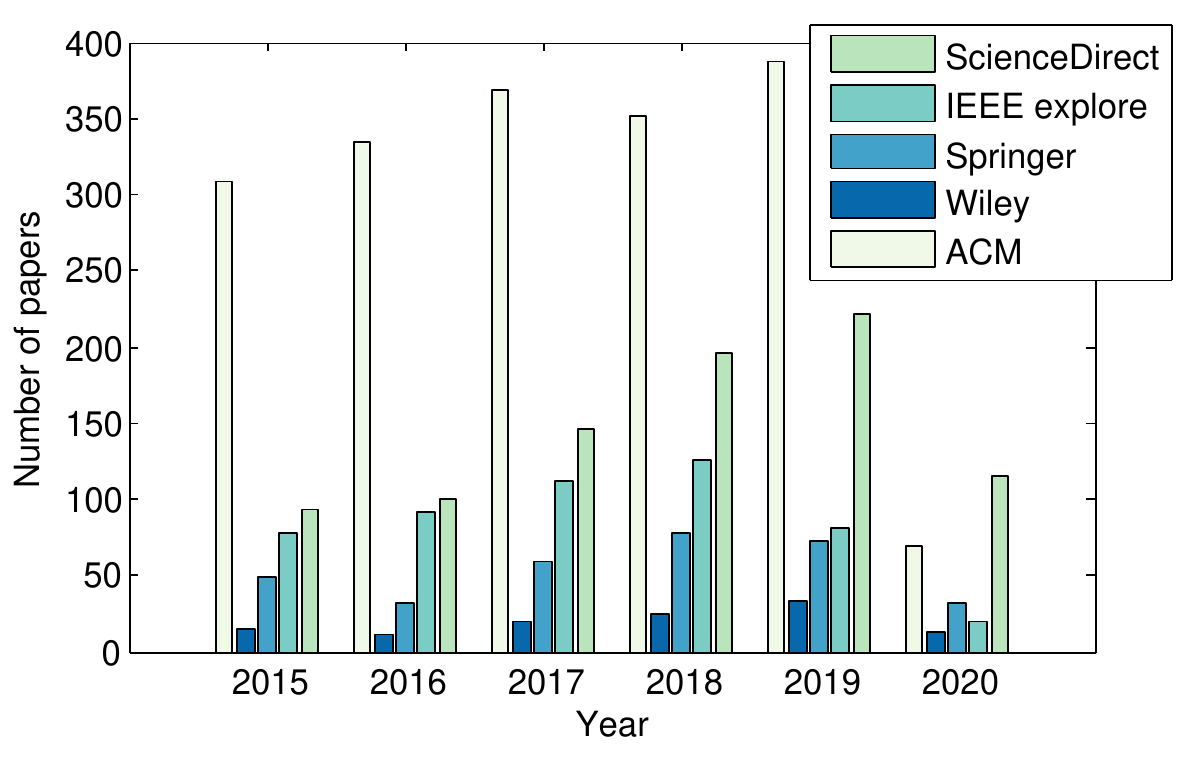}
    \caption{TBs in Smart Grid-based papers}
    \label{fig:2}
  \end{subfigure}
  \begin{subfigure}{0.32\textwidth}
    \centering
    \includegraphics[width=\linewidth]{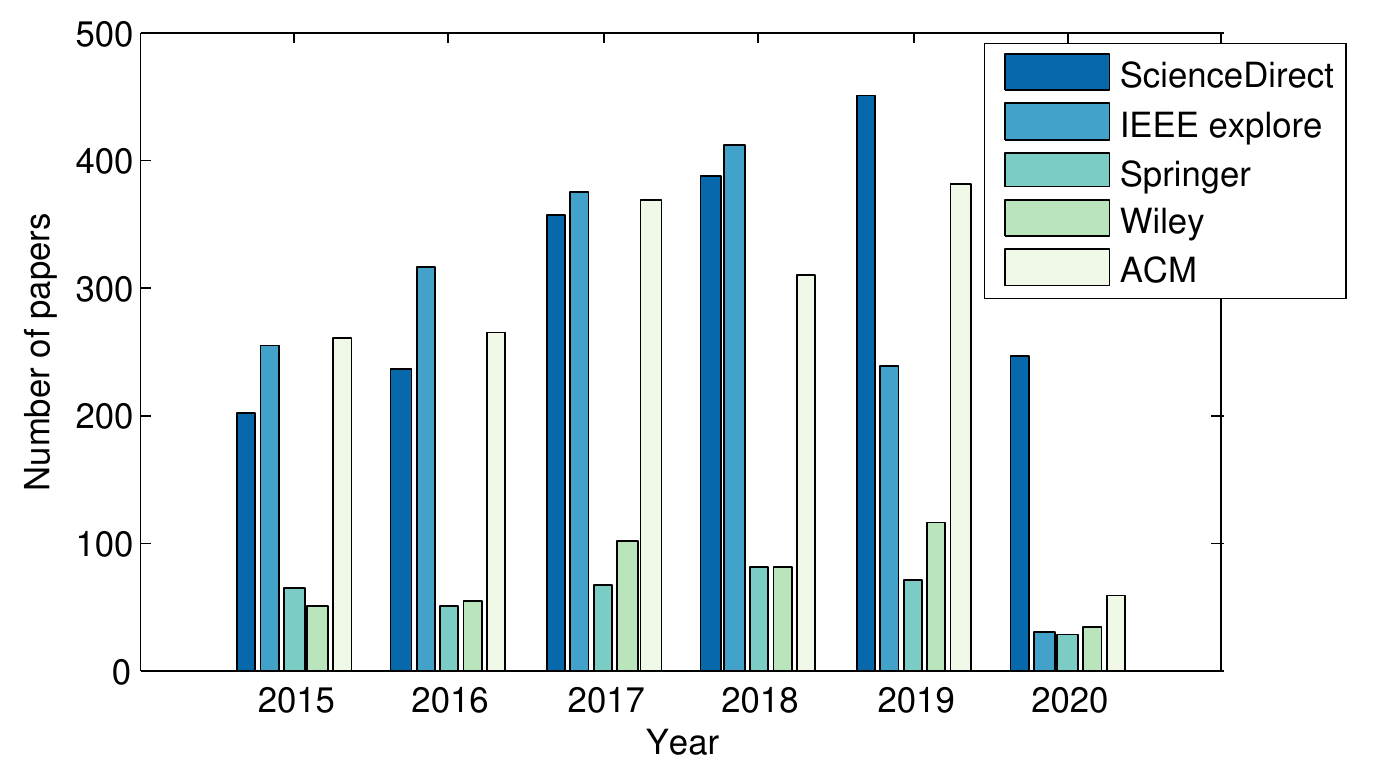}
    \caption{TBs in Cloud-based papers}
    \label{fig:3}
  \end{subfigure}

		\medskip

  \begin{subfigure}{0.32\textwidth}
    \centering
    \includegraphics[width=\linewidth]{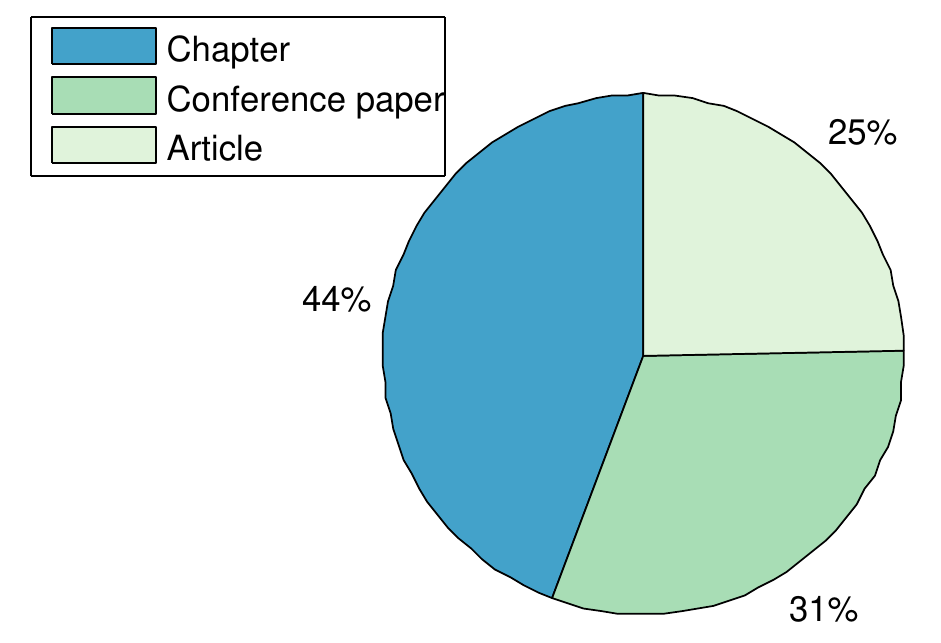}
    \caption{TBs in IoT type of papers}
    \label{fig:4}
  \end{subfigure}
  \begin{subfigure}{0.32\textwidth}
    \centering
    \includegraphics[width=\linewidth]{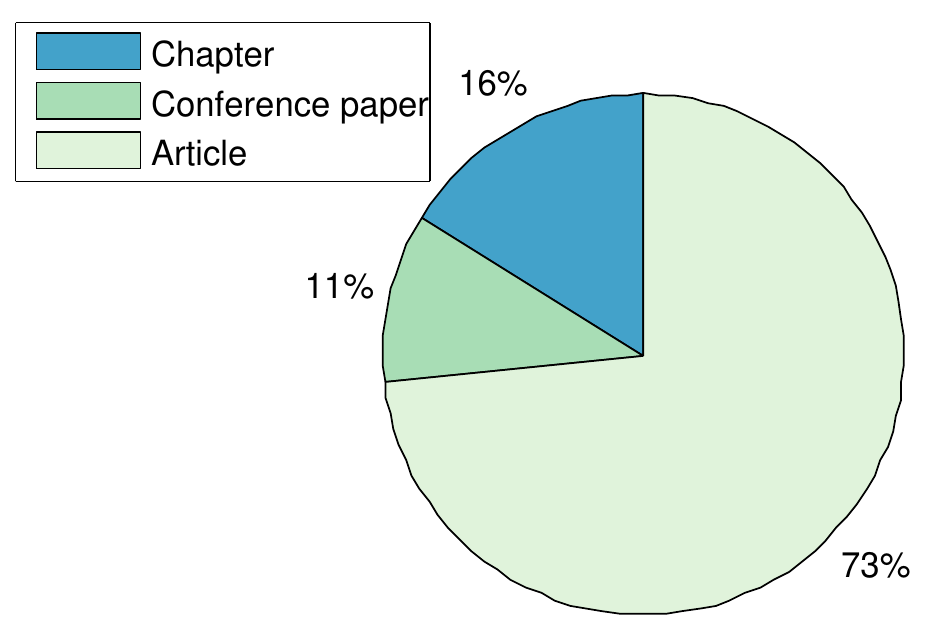}
    \caption{TBs in Smart Grids type of papers}
    \label{fig:5}
  \end{subfigure}
  \begin{subfigure}{0.32\textwidth}
    \centering
    \includegraphics[width=\linewidth]{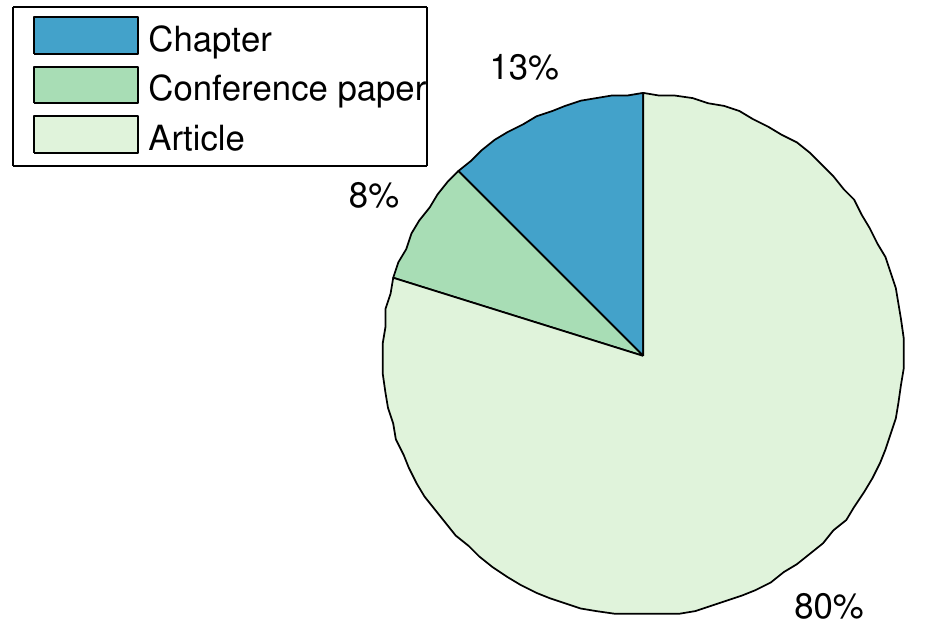}
    \caption{TBs in Cloud type of papers}
    \label{fig:6}
  \end{subfigure}
		
	\medskip
	
	  \begin{subfigure}{0.32\textwidth}
    \centering
    \includegraphics[width=\linewidth]{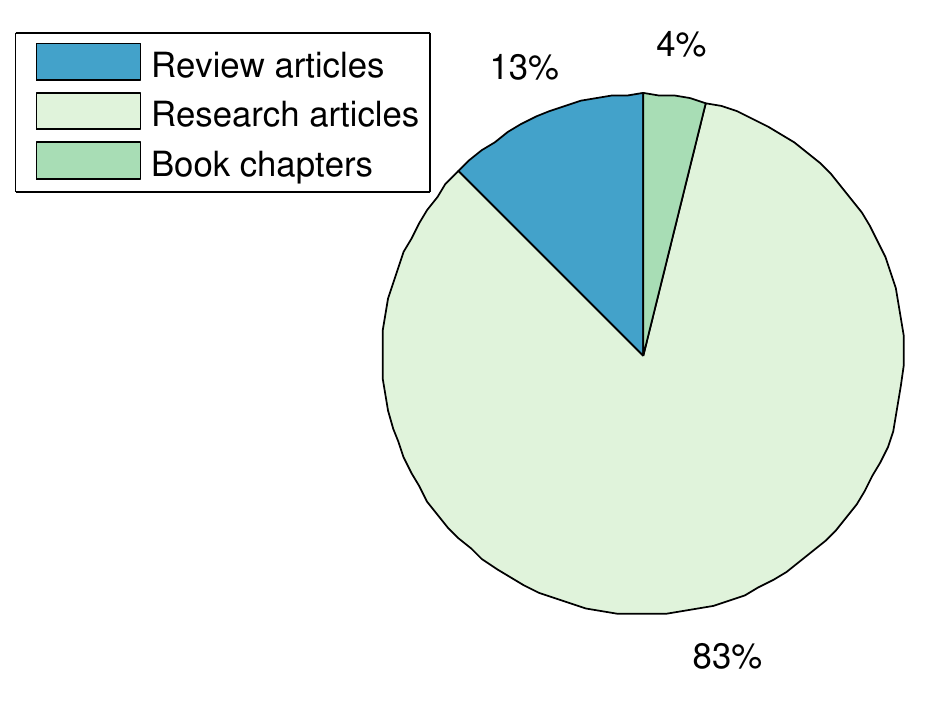}
    \caption{TBs in IoT type of papers}
    \label{fig:7}
  \end{subfigure}
  \begin{subfigure}{0.3\textwidth}
    \centering
    \includegraphics[width=\linewidth]{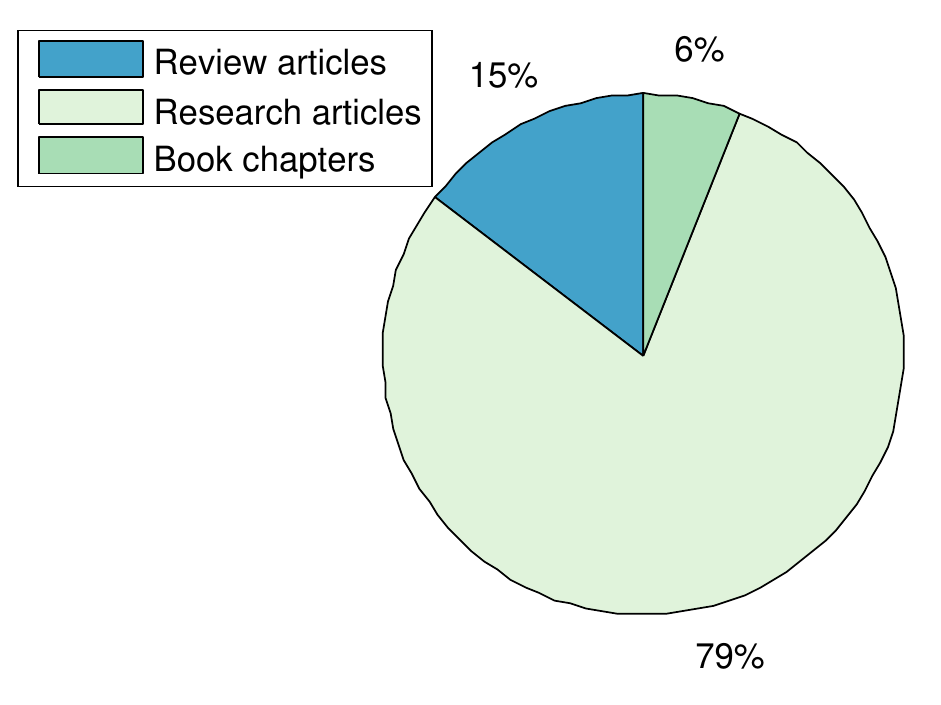}
    \caption{TBs in Smart grid type of papers}
    \label{fig:8}
  \end{subfigure}
  \begin{subfigure}{0.32\textwidth}
    \centering
    \includegraphics[width=\linewidth]{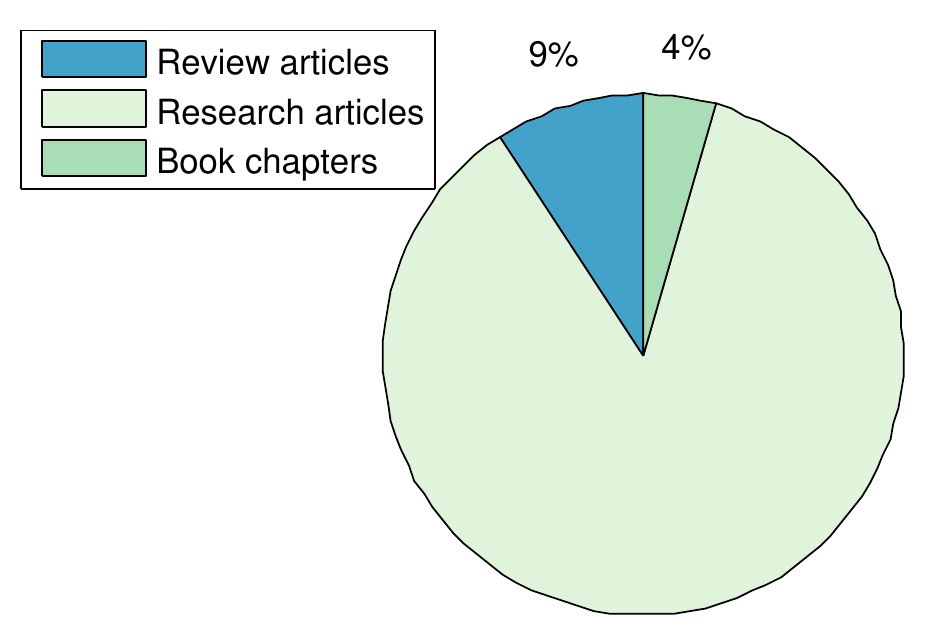}
    \caption{TBs in Cloud type of papers}
    \label{fig:9}
  \end{subfigure}
	
\caption{\textbf{Types of publications in TBs between 2015 and 2020 in Springer and ScienceDirect database respectively}}
  \label{fig:bift}

\end{figure}

\begin{figure*}[hbt!]
    \centering
    \begin{subfigure}{0.5\textwidth}
    \centering
        \includegraphics[scale=0.4]{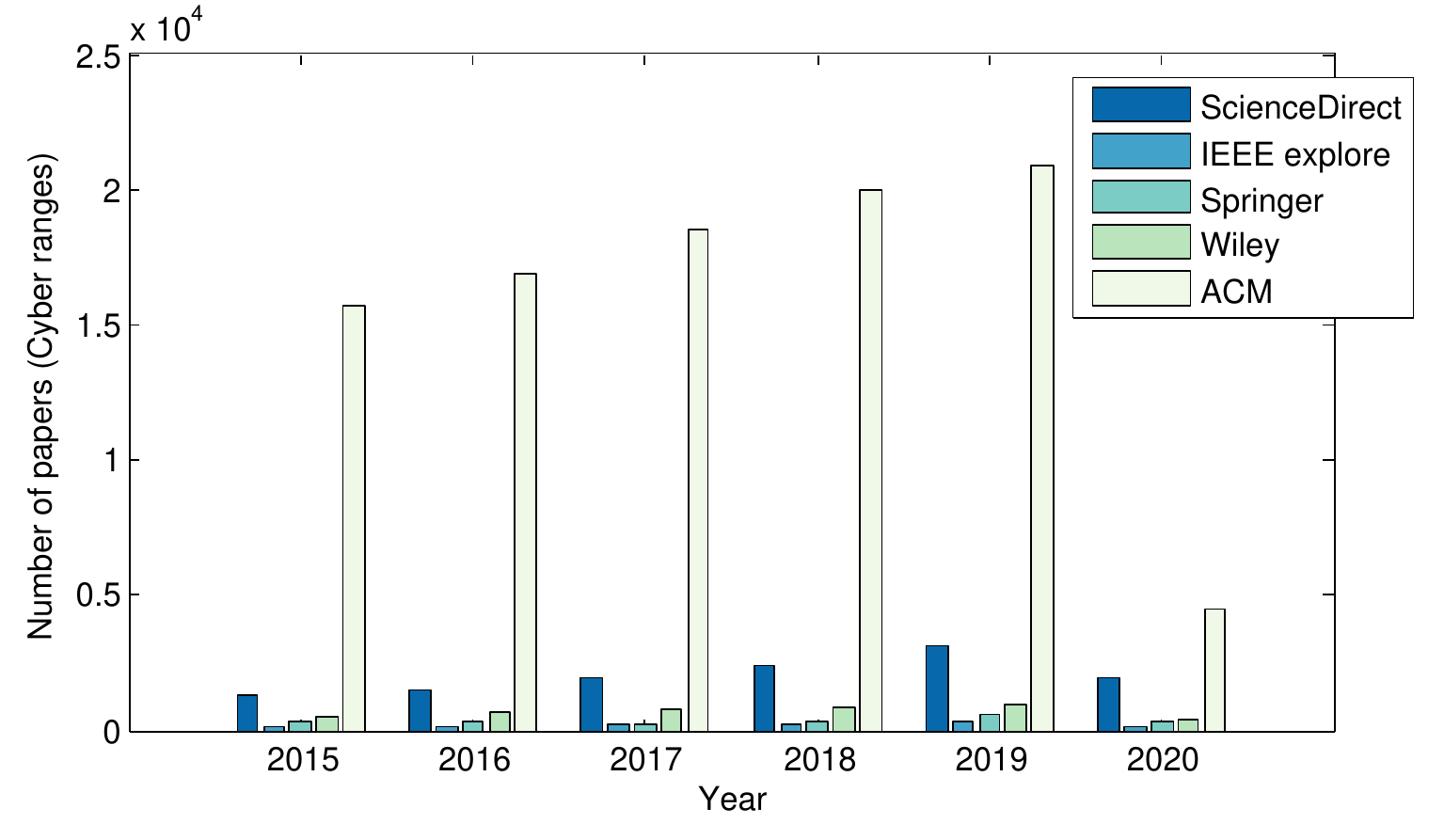}
				\caption{}
        \label{fig:first_sub}
				\end{subfigure}

    \begin{subfigure}{0.3\textwidth}
    
        \includegraphics[scale=0.5]{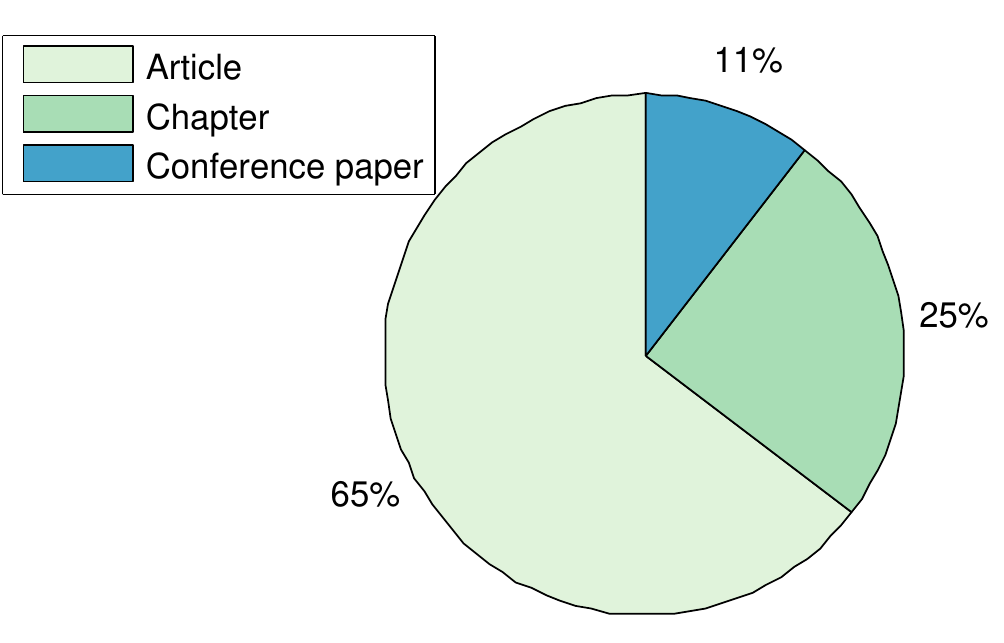}
				\caption{}
        \label{fig:85}
    
		\end{subfigure}
		\begin{subfigure}{0.3\textwidth}
    
        \includegraphics[scale=0.5]{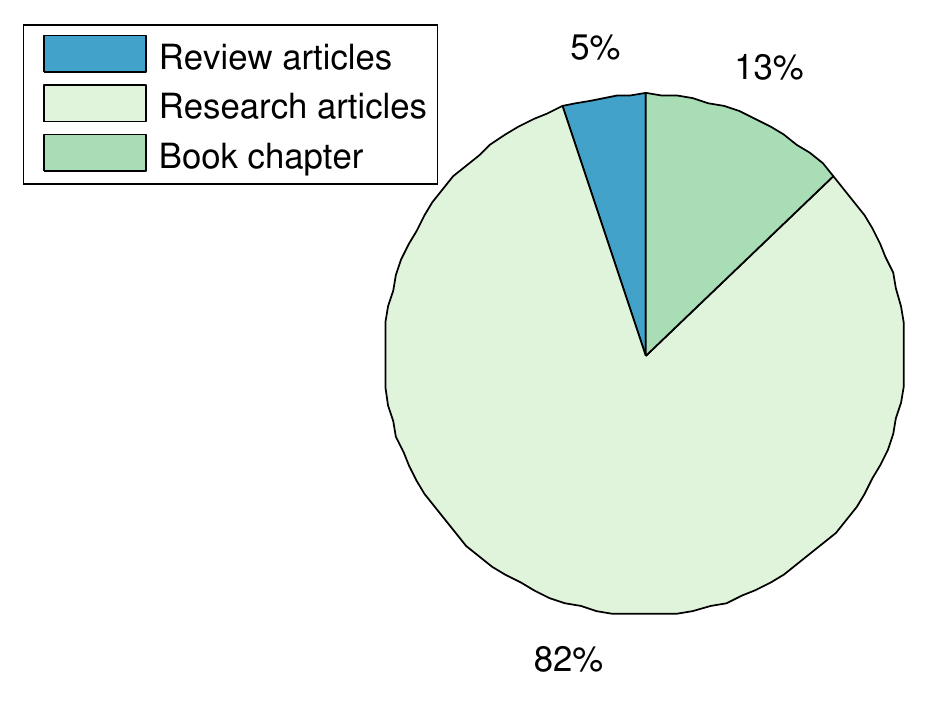}
				\caption{}
        \label{fig:86}
    
		\end{subfigure}
    \caption{\textbf{Evolution and classification of CRs publications}}
    \label{fig:sample_subfigures}
\end{figure*}

Figures \ref{fig:4}, \ref{fig:5},  \ref{fig:6}, \ref{fig:7}, \ref{fig:8}, \ref{fig:9}, \ref{fig:85} and \ref{fig:86} present the results of the review; Figures \ref{fig:4}, \ref{fig:5},  \ref{fig:6}, \ref{fig:7}, \ref{fig:8}, \ref{fig:9} summarise the publication types in Test-bed within the Springer and ScienceDirect databases. The percentage of papers on IoT Test-beds from the Springer database in Figure~\ref{fig:4} shows that 25\% of publications are journal articles, while 44\% are conference papers; 11\% of papers were published in conferences proceedings in Smart Grids (Figure~\ref{fig:5}); those published in the Cloud domain represent the highest percentage at (80\%) (Figure~\ref{fig:6}). 

Similarly, the review for CRs was restricted to the current prominent applications areas within  Military, Defence and Intelligence (Military), Education and Research (Education), and Industry and Commercial sectors (Industry). The data targeted focused on gaining insights on threat dynamics/proliferation, and emerging countermeasure strategies, a foundation for predicting future trends, technologies, and application areas. 

On inspection of Figure~\ref{fig:sample_subfigures}, it is evident that ACM publishes a greater number of papers in relation to CRs with progressive growth in number from 2015 through to 2019 (Figure~\ref{fig:first_sub}). Journal articles account for 64\% of publications, book chapters 25\% and conference proceedings 11\% (Figure~\ref{fig:85}). Also clear is that researched-based articles are more readily accepted for publication at a rate of 82\% compared to other types of articles such as Book Chapters at 13\% and Review articles at 5\% (Figure~\ref{fig:86}).

\begin{figure*}[hbt!]
\centering
\begin{tabular}{cccc}
\includegraphics[width=0.3\textwidth]{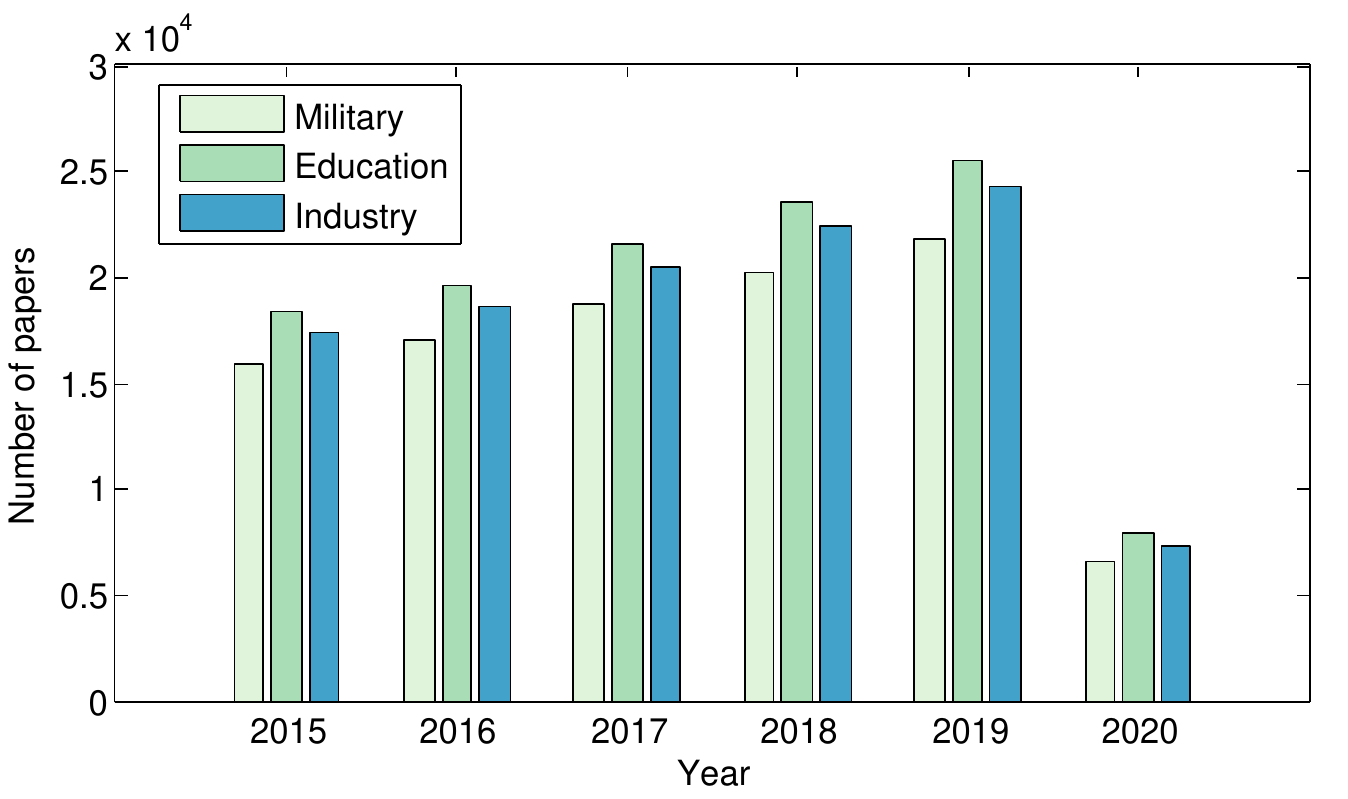} &
\includegraphics[width=0.3\textwidth]{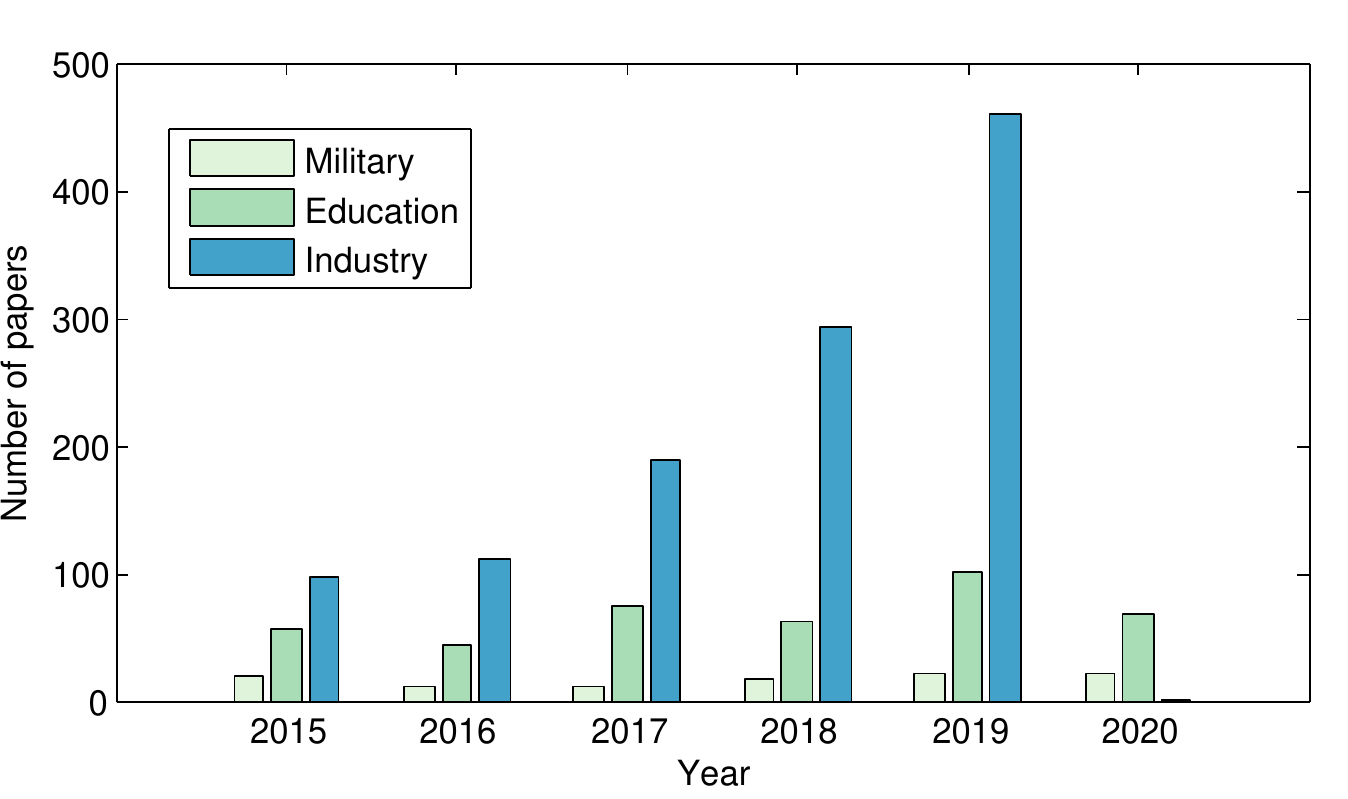} &
\includegraphics[width=0.3\textwidth]{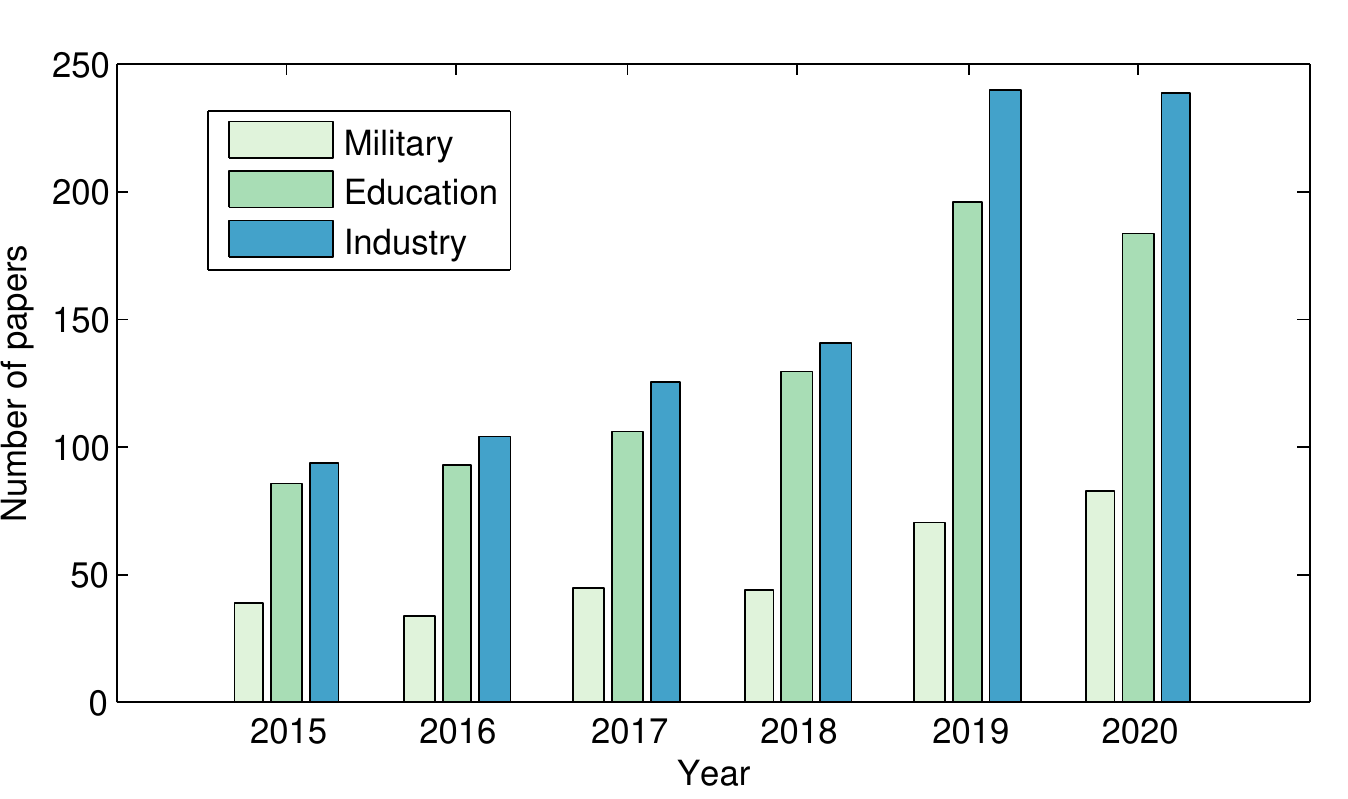} \\
\textbf{(a)}  & \textbf{(b)} & \textbf{(c)}  \\[6pt]
\end{tabular}
\begin{tabular}{cccc}
\includegraphics[width=0.3\textwidth]{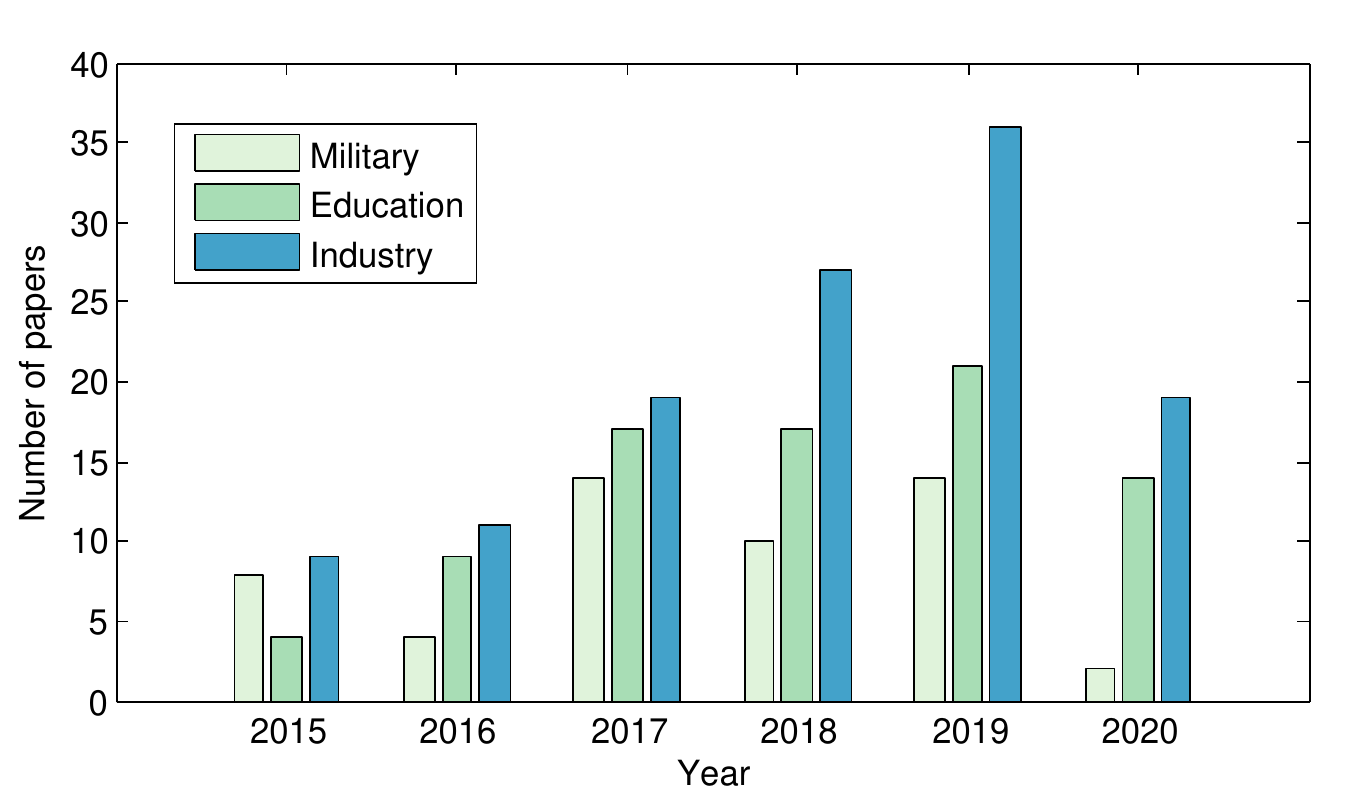} &
\includegraphics[width=0.3\textwidth]{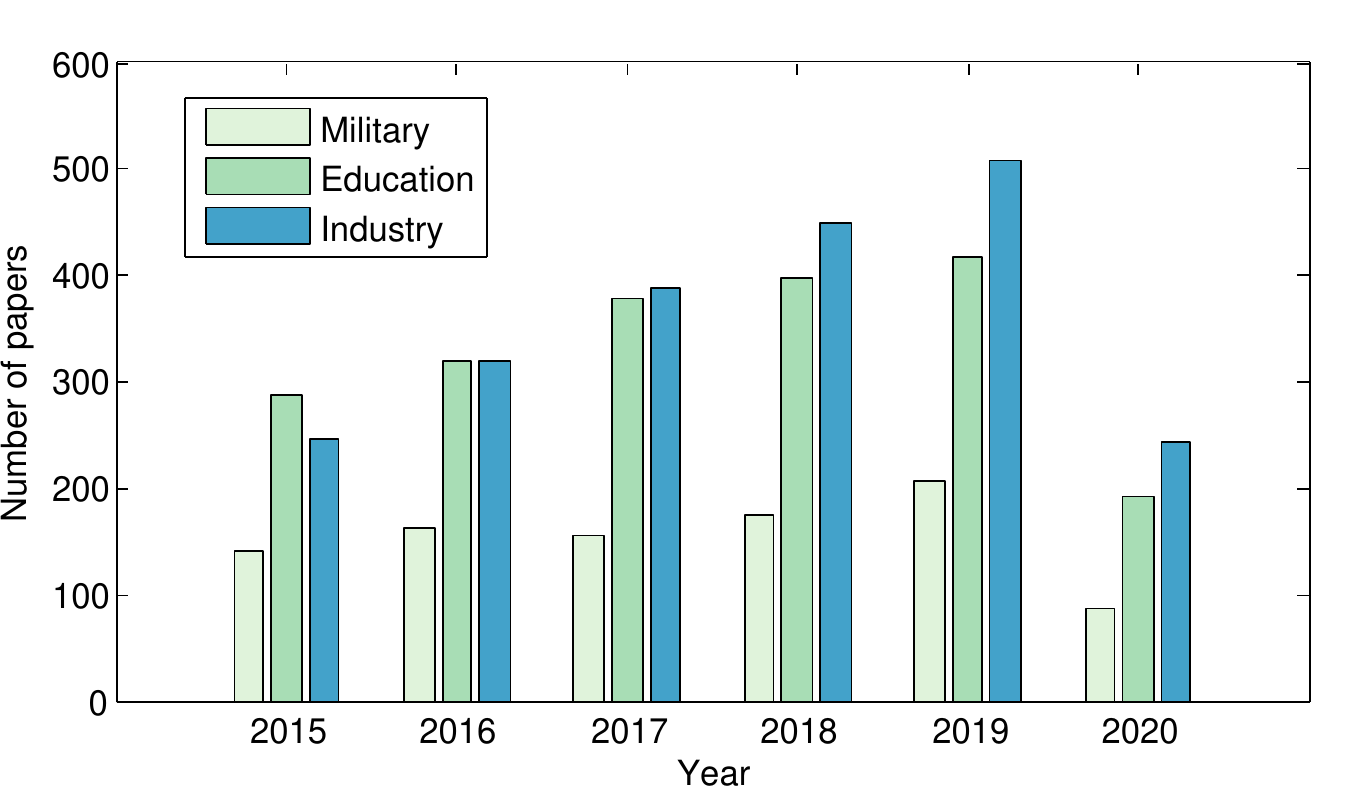} \\
\textbf{(d)}  & \textbf{(e)}  \\[6pt]
\end{tabular}
\caption{ \textbf{(a)} CR in ACM
\textbf{(b)} CR in ScienceDirect
\textbf{(c)} CR in Springer
\textbf{(d)} CR in IEEE
\textbf{(e)} CR in Willey}
\label{fig:Name}
\end{figure*}

 Further, ACM has published more Cyber-range related papers in the three domains of Military, Education, and Industry, followed by Wiley, ScienceDirect, Springer and IEEE  (Figure~\ref{fig:Name}). Cyber-range papers published in ACM are predominately in Education, followed by Industry and then Military. Wiley, ScienceDirect, Springer, and IEEE published more Industry-related Cyber-range papers than those in Education and Military.\\

\pagebreak
\section{\textbf{Related Work}}
\label{Section: RelatedWorks}
CR and TB solutions have been applied in both commercial and public settings such as the military/defence, intelligence, education, research, and training~\cite{davis2013survey}. The extensive usage is driven by the need to be cyber-security prepared against persistent threats to critical infrastructures and businesses. The first publicly reported CR was the National Cyber-Range~(NCR)~\cite{ranka2011national}, created by the US Department of Defence. However, as described in~\cite{ranka2011national}, other `stealthy' CRs and TBs existed across the world for cyber-warfare training in advance of the NCR. 

Leblanc~\textit{et al.}~\cite{leblanc2011overview} surveyed the state-of-the-art of 13 simulation-centric CRs categorising them into private, academic or public sector research. The review, while useful, dates back to 2011 and given the significant advances in the functionality of technologies and complexity of threat dynamics limits the value of their findings in the goal of predicting the future evolution. 

Davis and Magrath in~\cite{davis2013survey} conducted a survey of CRs in the public domain focusing on 30 existing systems in 2013. The review focused primarily on the merits of each approach in respect of their functionalities with emphasis on cost-effectiveness. The classification was segmented in terms of military and government; academic; and commercial and further categorised as either simulation or emulation-driven implementations. The authors concluded that emulation-driven CRs utilised TBs and were proven to be effective environments for training and test. The trade-off between highly functional, robust training environments and the concomitant cost implications as a result of the complexity of the implementations was stressed. Inherent within the trade-off, is the provision for the sharing of resources and/or virtualisation. Conversely, simulation-based CRs are implemented solely through software that model real world scenarios, and are thus easily scalable. However, emulation-driven CRs can be validated more readily for performance~\cite{davis2013survey}. 

As the review~\cite{davis2013survey} was carried out considering CRs and TBs before 2013, the conclusions on the trade-off between functionalities and cost-effectiveness has limited value in the determination of the future evolution of the platforms.

Priyadarshini~\cite{priyadarshini2018features} also reported the results of a review on CRs in 2018, culminating in the definition of the features and capabilities of an `Ideal CR'; the components, scenarios, and capabilities of the CR at the University of Delaware~(CRUD) were used as the foundation for the definition of the future platform. The bench-marking did not consider the needs of the applications viz. to facilitate training, education, and research addressing recent and future threat profiles, their proliferation and modes of attack.

The most recent literature by Yamin~\textit{et al.}~\cite{yamin2019cyber} reviews unclassified CRs and security TBs. The authors propose a taxonomy with reference to the architecture, scenarios, capabilities, roles, and tools as the criteria. The main output is a proposed baseline to aid the development and evaluation of CRs. 

The above reviews provided valuable insights into CR/TB technologies with potential to facilitate training in the management of persistent cyber-threats, their changes in perspective, execution, and patterns. However, as a consequence of the dynamic and rapid development of technologies and the enhanced capabilities they provide, the conclusions are limited in the goal of predicting the evolution of future CR/TB platform capabilities and the scope of training they support. The taxonomy that captures these dynamic trends needs to be re-established in the light of advances made in the recent past.

\section{\textbf{Systematic Review}}
A systematic review to bring to fore the context of this paper is presented using this section.
\label{Section: SystematicReview}
\subsection{\textbf{Cyber-Ranges and Test-Beds}}
Table~\ref{tab:SummaryTable} is a summary of CRs and TBs covering a period of five years, the basis for a systematic review to predict the threat landscape, dimension and proliferation taking into consideration continual technological advancement. The Table is segmented into a number of categories viz. Military, Defense and Intelligence~(MDI), Academic~(Aca), Enterprise and Commercial~(EC), Service Providers~(SP), Open Source~(OS), Law Enforcement~(LE), Government~(Gvt), Mode of Deployment~(Deploy), Area of Specialties~(Specialty), Types~(Type), The Team it supports~(Team), The Testing Environment~(TE) and Method of Experimentation~(ME).

\begin{footnotesize}

\begin{longtable}{|m{0.12\linewidth}|C{0.03\linewidth}|C{0.025\linewidth}|C{0.025\linewidth}|C{0.025\linewidth}|C{0.025\linewidth}|C{0.025\linewidth}|C{0.025\linewidth}|C{0.025\linewidth}|C{0.05\linewidth}|C{0.06\linewidth}|C{0.04\linewidth}|C{0.03\linewidth}|C{0.025\linewidth}|C{0.025\linewidth}|}

\caption{Summary Table of Related Works} \label{tab:SummaryTable} \\

\hline
Categories & Ref & \rotatebox{90}{MDI} & \rotatebox{90}{Aca} & EC & SP & OS & LE & \rotatebox{90}{Govt} & \rotatebox{90}{Deploy} & \rotatebox{90}{Specialty} & Type & \rotatebox{90}{Team} & TE &ME\\ \hline
\endfirsthead

\hline
Categories & Ref & \rotatebox{90}{MDI} & \rotatebox{90}{Aca} & EC & SP & OS & LE & \rotatebox{90}{Govt} & \rotatebox{90}{Deploy} & \rotatebox{90}{Specialty} & Type & \rotatebox{90}{Team} & TE &ME\\ \hline
\hline
\endhead

\hline \multicolumn{15}{|r|}{{Continued \ldots}} \\ \hline
\endfoot

\hline \hline
\endlastfoot

NCR	&\cite{ranka2011national} \cite{davis2013survey} \cite{priyadarshini2018features} &	Y&	N&	N&	N&	N&	N&	Y&	C,  VPN &	ST,  NS&	Fed&	R, B, Gy&	VM&	E\\
\hline
Virginia CR	&\cite{radziwill2017virginia} \cite{priyadarshini2018features}	&N&	Y&	N&	N&	N&	N&	N&	C&	ST&	Pub,  Prv&	R, B	&VM&	N\\
\hline
Michigan CR&\cite{priyadarshini2018features}&	N&	Y&	N&	N&	N&	Y&	N&	C, VPN&	ST,  AS,  CSE&	Pub,  Prv&	R,  B	&VM,   SB&	N\\
\hline
Pinecone CR	&\cite{priyadarshini2018features}&	N&	N&	Y&	N&	N&	N&	N&	N&	N&	N&	N&	N&	N\\
\hline
IBM X-Force	&\cite{priyadarshini2018features}&	N&	N&	Y&	N&	N&	N&	N&	C&	N&	Prv&	R, B&	VM,  SB&	N\\
\hline
Cyberbit CR	&\cite{priyadarshini2018features}&	N&	N&	N&	Y&	N&	N&	N&	N&	N&	N&	N&	N&	N\\
\hline
Arizona CWR	&\cite{priyadarshini2018features}&	N&	N&	N&	N&	Y&	N&	N&	N&	CSE&	N&	N&	N&	N\\
\hline
CRATE&	\cite{sommestad2015experimentation} \cite{priyadarshini2018features}&	Y&	N&	N&	N&	N&	N&	N&	C,  VPN&	N	&Fed&	R,  B&	VM&	N\\
\hline
Cisco CR&\cite{priyadarshini2018features}&	N&	Y&	Y&	N&	N&	N&	N&	C,  VPN&	N&	Pub,  Prv&	R,  B,  Gn&	VM&	N\\
\hline
NATO CR	&\cite{pernik2014improving} \cite{priyadarshini2018features}&	N&	Y&	N&	N&	N&	N&	N&	C,  VPN&	N&	Fed&	R,  B,  G,  W,  Y&	VM,   SB&	N\\
\hline
DoD CR&\cite{ferguson2014national} \cite{priyadarshini2018features}&	Y&	N&	N&	N&	N&	N&	N&	C,  VPN&	N	&Fed&	R,  B&	VM,   SB&	N\\
\hline
Raytheon CR	&\cite{priyadarshini2018features}&	N&	N&	Y&	N&	N&	N&	N&	C,  VPN&	N&	Fed&	R,  B&	N&	N\\
\hline
Baltimore CR&	\cite{priyadarshini2018features}&	N&	N&	Y&	N&	N&	N&	Y&	C,  VPN&	N&	Pub,  Prv	&R,  B	&N&	N\\
\hline
Florida CR	&\cite{priyadarshini2018features}&	Y&	Y&	N&	N&	N&	N&	Y&	C&	PT,  EH, NS,  SS&	Fed, Pub, Prv&	R, B&	N&	N\\
\hline
CRUD	&\cite{priyadarshini2018features}&	N&	Y&	N&	N&	N&	N&	N&	N&	N&	N&	R, B, P&	VM&	N\\
\hline
Regent CR&	\cite{priyadarshini2018features}&	Y&	Y&	Y&	N&	N&	N&	N&	N&	RA, M, TV, DF&	N&	N&	N&	N\\
\hline
Wayne CR&\cite{priyadarshini2018features}&	N&	Y&	Y&	N&	N&	N&	N&	C&	EH, CTF, PT, EH&	N&	N&	SB&	N\\
\hline
Arkansas CR	&\cite{priyadarshini2018features}&	N&	Y&	N&	N&	N&	N&	N&	C&	PT&	N&	N&	VM&	N\\
\hline
Georgia CR&\cite{priyadarshini2018features}&	N&	N&	Y&	N&	N&	N&	Y&	N&	N&	N&	N&	N&	N\\
\hline
SIMTEX	&\cite{davis2013survey}&	Y&	N&	N&	N&	N&	N&	Y&	N&	CSE&	Pub&	N&	N&	S\\
\hline
CAAJED&\cite{mudge2008cyber} \cite{davis2013survey}&	Y&	N&	N&	N&	N&	N&	N&	N&	CSE&	Pub&	N&	N&	S\\
\hline
SAST&\cite{meitzler2009security} \cite{davis2013survey}&	Y&	N&	N&	N&	N&	N&	N&	N&	CSE&	Pub&	N	&N&	S\\
\hline
StealthNet&\cite{varshney2011live} \cite{davis2013survey}&	Y&	N&	N&	N&	N&	N&	N&	N&	CSE&	Pub&	N&	N&	S\\
\hline
SECUSIM&\cite{chi2003role} \cite{davis2013survey}&	N&	Y&	N&	N&	N&	N&	N&	N&	CSE&	Pub&	N	&N	&S\\
\hline
RINSE&\cite{liljenstam2005rinse} \cite{davis2013survey}&	N&	Y&	N&	N&N&	N&	N&	N&	CSE&	Pub&	N	&N	&S\\
\hline
NetENGINE&\cite{brown2003simulation} \cite{davis2013survey}&	N&	Y&	N&	N&	N&	N&	Y&	N&	CSE&	Prv&	N&	N&	S\\
\hline
ARENA&\cite{kuhl2007cyber} \cite{davis2013survey}&	N&	Y&	N&	N&	N&	N&	N&	N&	CSE&	Pub&	N	&N&	S\\
\hline
OPNET-based&	\cite{zhou2003frequency} \cite{davis2013survey}&	N&	N&	N&	N&	N&	N&	N&	N&	NA&	Pub&	N&	N&	N\\
\hline
LARIAT& \cite{rossey2002lariat} \cite{davis2013survey}&	Y&	N&	N&	N&	N&	N&	N&	N&	CSE&	Pub&	N	&N&	S\\
\hline
VCSTC& \cite{pederson2008virtual} \cite{davis2013survey}&	N&	Y&	N&	N&	N&	N&	Y&	N&	CSE	&Pub&	N	&N	&S\\
\hline
Breaking Point&\cite{davis2013survey}&	N&	N&	Y&	N&	N&	N&	N&	N&	Training&	Prv&	N&	N&	S\\
\hline
Exata	&\cite{davis2013survey}&	Y&	N&	Y&	N&	N&	N&	N&	N&	Training&	Prv	&N&	N&	S\\
\hline
PlanetLab&\cite{davis2013survey}&	N&	N&	N&	N&	N&	N&	N&	N&	Training, CSE&	Fed, Pub&	N	&N&	O\\
\hline
X-Bone&\cite{davis2013survey}&	N&	N&	N&	N&	N&	N&	N&	N&	CSE&	Prv&	N&	N&	O\\
\hline
JIOR&	\cite{davis2013survey}&	Y&	N&	N&	N&	N&	N&	Y&	N&	CSE, Training&	Fed&	N&	N&	E\\
\hline
INL&\cite{anderson2009cyber} \cite{davis2013survey}&	Y&	N&	N&	N&	N&	N&	Y&	N&	CSE, Training&	Fed&	N&	N&	E\\
\hline
Emulab&\cite{siaterlis2012use} \cite{davis2013survey}&	N&	Y&	N&	N&	N&	N&	N&	N&	CSE, Research&	Fed&	N&	N&	E\\
\hline
DETER	&\cite{davis2013survey}&	N&	Y&	N&	N&	N&	N&	N&	N&	CSE, Research&	Fed&	N&	N&	E\\
\hline
Virtualised CR&\cite{mayo2009approaches} \cite{davis2013survey}&	N&	Y&	N&	N&	N&	N&	N&	N&	CSE, Research&	Fed&	N&	N&	E\\
\hline
Reassure &\cite{thomas2009mandatory} \cite{davis2013survey}&	N&	Y&	N&	N&	N&	N&	N&	N&	Research&	Pub&	N&	N&	E\\
\hline
Northrop G&\cite{davis2013survey}	&Y&	N&	Y&	N&	N&	N&	N&	N	&CSE, Training&	Prv&	N&	N&	E\\
\hline
Counter HC&\cite{davis2013survey}&	N&	N&	Y&	N&	N&	N&	N&	N&	Training, CSE&	Prv&	N&	N&	NA\\
\hline
Detica	&\cite{davis2013survey}&	N&	N&	Y&	N&	N&	N&	N&	N&	Training&	Prv&	N&	N&	N\\
\hline
ATC	&\cite{brueckner2008automated} \cite{davis2013survey}&	N&	N&	Y&	N&	N&	N&	N&	N&	Training&	Prv&	N&	N&	LS\\
\hline
Testbed@ TWISC &\cite{tsai2018testbed} \cite{yamin2019cyber} &	N&	Y&	N&	N&	N&	N&	N&	C&	Research,  ST&	Prv&	N&SB	&E\\
\hline
INSALATA &\cite{herold2017achieving} \cite{yamin2019cyber} &	N&	N&	N&	N&	Y&	N&	N&	C&	NS, Research&	Prv&	N & \rotatebox{90}{Hybrid}	&E\\
\hline
CyberVan &\cite{chadha2016cybervan} \cite{yamin2019cyber} &	Y&	N&	N&	N&	N&	N&	N&	C&ST, NT&	Pub&N&\rotatebox{90}{Hybrid}	&S\\
\hline
SoftGrid &\cite{gunathilaka2016softgrid} \cite{yamin2019cyber} &	Y&	N&	N&	N&	Y&	N&	N&	C&ST, NT, AS&	Prv&N&H	&E\\
\hline

\multicolumn{15}{l}{\textbf{Legend:}} \\ 
\multicolumn{4}{l}{Aca: Academic or Research} & \multicolumn{6}{l}{H: Hardware} & \multicolumn{5}{l}{RA: Ransomware Attacks} \\ 
\multicolumn{4}{l}{AS: Attack Simulation} & \multicolumn{6}{l}{LE: Law Enforcement} & \multicolumn{5}{l}{S: Simulation} \\ 
\multicolumn{4}{l}{B: Blue} & \multicolumn{6}{l}{M: Monitoring} & \multicolumn{5}{l}{SB: Sandbox} \\ 
\multicolumn{4}{l}{C: Cloud} & \multicolumn{6}{l}{MDI: Military, Defense and Intelligence} & \multicolumn{5}{l}{SP: Service Provider} \\ 
\multicolumn{4}{l}{CSE: Cyber Security Exercise} & \multicolumn{6}{l}{ME: Method of Experimentation} & \multicolumn{5}{l}{SS: System Security} \\ 
\multicolumn{4}{l}{CTF: Catch-The-Flag} & \multicolumn{6}{l}{N: No/Not Available} & \multicolumn{5}{l}{ST: Security/Software Testing} \\ 
\multicolumn{4}{l}{DF: Digital Forensic} & \multicolumn{6}{l}{NS: Network Security} & \multicolumn{5}{l}{TE: Testing Environment} \\ 
\multicolumn{4}{l}{E: Emulation} & \multicolumn{6}{l}{O: Overlay} & \multicolumn{5}{l}{TV: Threat and Vulnerability} \\ 
\multicolumn{4}{l}{EC: Enterprise and Commercial} & \multicolumn{6}{l}{OS: Open Source} & \multicolumn{5}{l}{VCN: Virtual Clone Network} \\ 
\multicolumn{4}{l}{EH: Ethical hacking} & \multicolumn{6}{l}{P: Purple} & \multicolumn{5}{l}{VM: Virtual Machine} \\ 
\multicolumn{4}{l}{Fed: Federated} & \multicolumn{6}{l}{Prv: Private} & \multicolumn{5}{l}{VPN: Virtual Private Network} \\ 
\multicolumn{4}{l}{Gn: Green} & \multicolumn{6}{l}{PT: Penetration Testing} & \multicolumn{5}{l}{W: White} \\ 
\multicolumn{4}{l}{Govt: Government} & \multicolumn{6}{l}{Pub: Public} & \multicolumn{5}{l}{Y: Yellow} \\ 
\multicolumn{4}{l}{Gy: Grey} & \multicolumn{6}{l}{R: Red} & \multicolumn{5}{l}{Y: Yes/Available} \\ 

\end{longtable}

\end{footnotesize}

\begin{itemize}
    \item \textbf{Application Domains:}
A total of 44 CRs were categorised, including the CRs surveyed by~\cite{davis2013survey}~and~\cite{priyadarshini2018features}. Figure~\ref{fig:CRDomain} shows that CRs have been predominantly used for academic purposes in education and research at 31\%. The result differs from that of Davis and Magrath~\cite{davis2013survey} of 2013, where the predominant use of CRs was in the training for cyber-security, a paradigm shift in the main application. The trend is also consistent with the findings presented in Section \ref{Section: Methodology}; that the bulk of CR papers were published by the academic community reporting on applications in teaching, learning, and research; followed by Enterprises and Commercial organisations, as well as, Military Defence and Intelligence for training purposes such as cyber-defence preparedness both at 24\% respectively. The use in of CRs in Government was at 15\% rate, while other application areas such as Law Enforcement, Service Providers, and Open-Source constitute only 2\% of the manuscripts surveyed.

\begin{figure}[H]
    \centering
    \includegraphics[width=6cm]{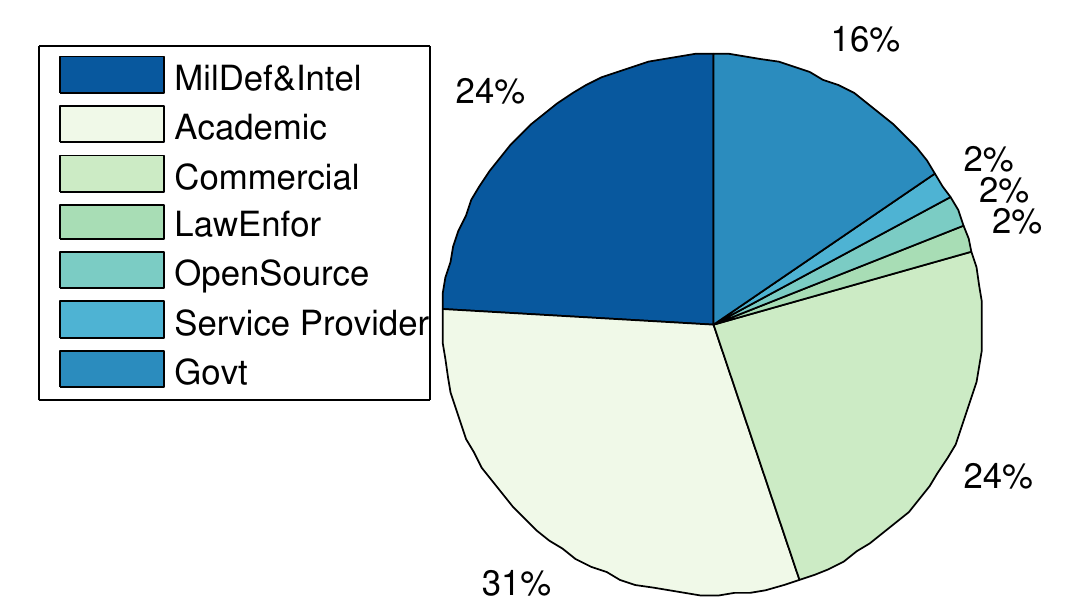}
    \caption{\textbf{Cyber-Range Domain of Applications}}
    \label{fig:CRDomain}
\end{figure}

Only five TBs were identified (Table~\ref{tab:SummaryTable}), of which three were applied in academia for the purposes of education and research; 
Testbed@TWISC~\cite{tsai2018testbed}, CyberVan~\cite{chadha2016cybervan}, and INSALATA~\cite{herold2017achieving}. SoftGrid~\cite{gunathilaka2016softgrid} and systems such as LARIAT~\cite{haines2001extending}~\cite{rossey2002lariat}, have been applied in defence and intelligence training. \\

\item\textbf{Types:}
Figure~\ref{fig:CRTypes} shows that public and federated CRs are predominant in use at 30\% respectively, private at 24\%, a combination of Public-Private at 11\%, a combination of Federated-Public-Private at 3\% and Federated-Public at 2\%. A link between the cyber-security preparedness application with the type of technology is evident. The predominant domain of application is for academic purposes and the institutions that provide education and research are mostly public with international collaborative perspectives, thereby suggesting an inter-relationship.

\begin{figure}[H]
    \centering
    \includegraphics[width=6cm]{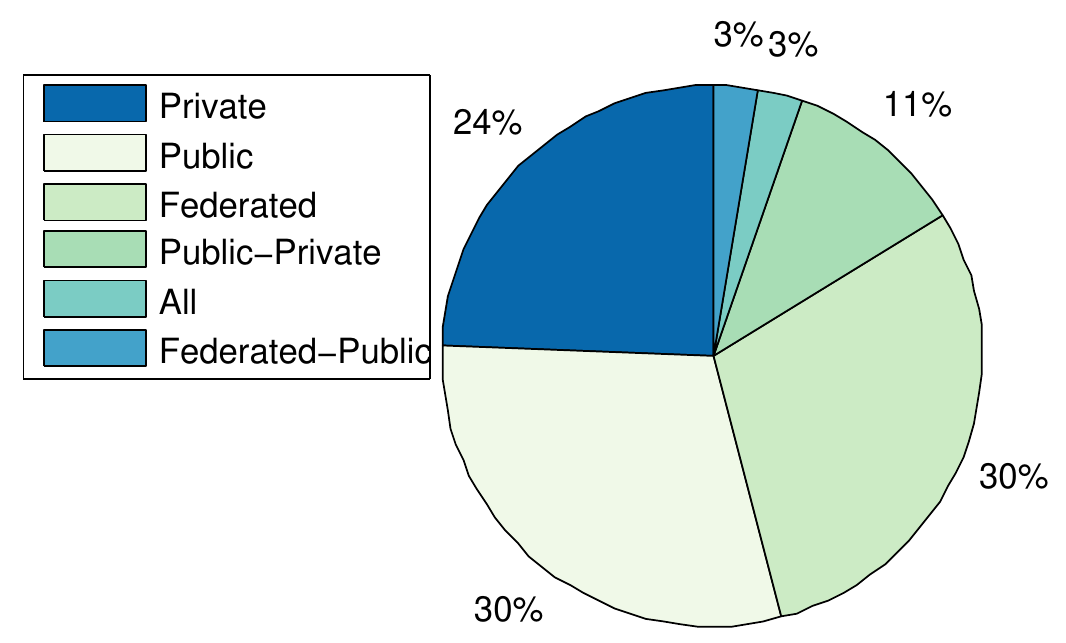}
    \caption{\textbf{Cyber-Range Types}}
    \label{fig:CRTypes}
\end{figure}

CyberVan~\cite{chadha2016cybervan} is a public type of Test-bed, while Testbed@TWISC~\cite{tsai2018testbed}, INSALATA~\cite{herold2017achieving}, SoftGrid~\cite{gunathilaka2016softgrid} and LARIAT~\cite{haines2001extending}~\cite{rossey2002lariat} are private TBs. \\ 

\item\textbf{Team Formation:}
Team formations are central to training through exercises emulating operations. Teams are formed depending on the type of exercise; (1)~Red team acts as adversaries by launching attacks on the network system; (2)~Blue team is responsible for defending against an adversary attack; (3)~White team for administrative management; (4)~Purple team sets objectives for offensive and defensive strategies; (5)~Green team is responsible for maintaining network efficiency; (6)~Grey team conducts non-malicious activity; and (7)~Yellow team acts as a motivator during each exercise. From the survey Red-Blue team formation is most prominent at 67\%, an indication that many CRs are dedicated to cyber-attack and defence training and exercises, followed by Red-Blue-Grey teams at 9\% and others such as Red-Blue-Green, Red-Blue-Green-White-Yellow and Red-Blue-Purple with 8\% each as shown in Figure~\ref{fig:CRTeams}. The training of teams on operational environments is restrictive in the scope of threat conditions that can be established as it compromises business continuity.

\begin{figure}[H]
    \centering
    \includegraphics[width=6cm]{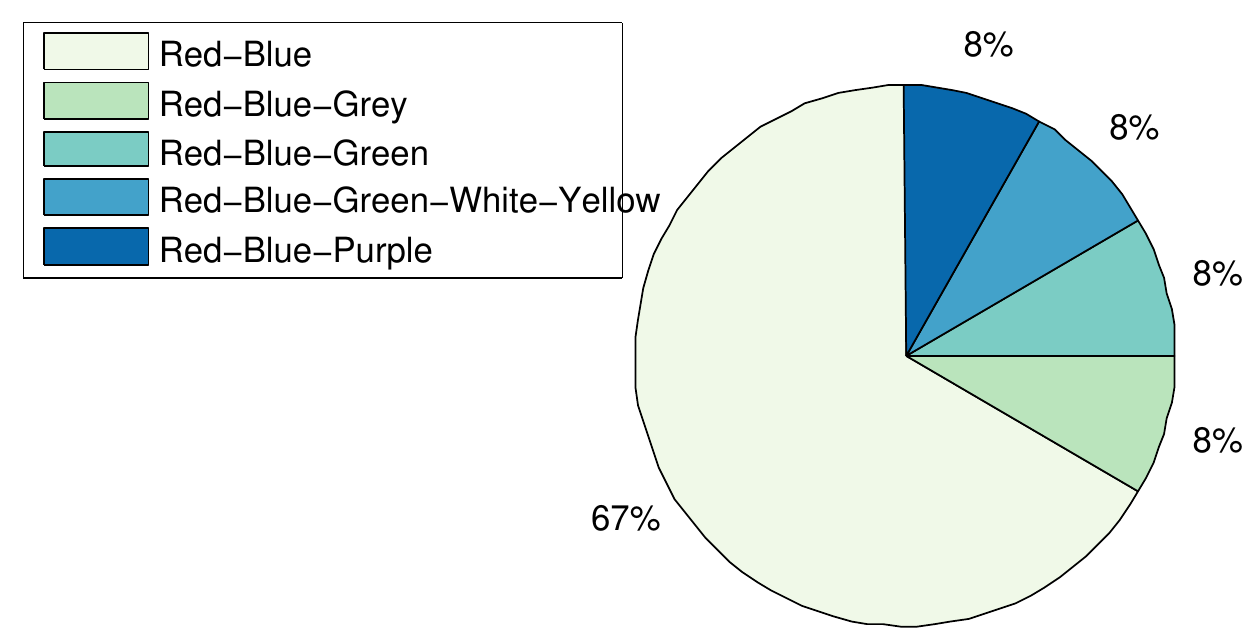}
    \caption{\textbf{Cyber-Range Teams}}
    \label{fig:CRTeams}
\end{figure}

\item\textbf{Methods of Experimentation:}
Figure~\ref{fig:CRMethods} highlights that simulation is the most common implementation methodology at 60\%, followed by emulation at 38\%, overlay at 8\%, and finally live scenario demonstrations at 4\%. 

\begin{figure}[H]
    \centering
    \includegraphics[width=5.6cm]{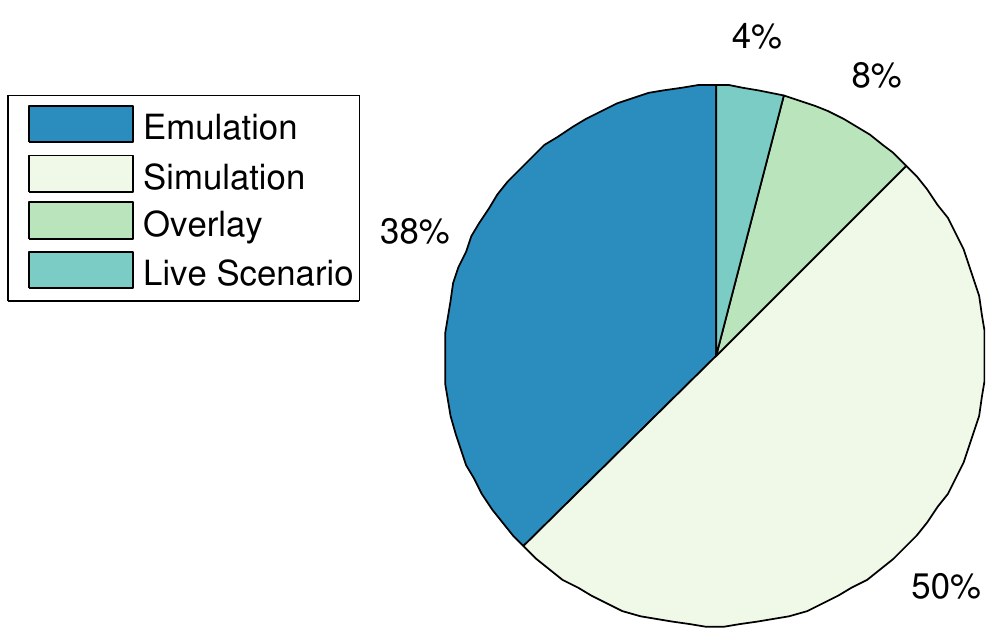}
    \caption{\textbf{Cyber-Range Methods of Experimentation}}
    \label{fig:CRMethods}
\end{figure}

 Testbed@TWISC~\cite{tsai2018testbed}, INSALATA~\cite{herold2017achieving}, SoftGrid~\cite{gunathilaka2016softgrid} and LARIAT~\cite{haines2001extending}~\cite{rossey2002lariat} all use emulation techniques except CyberVan~\cite{chadha2016cybervan} that is based on simulation.
\end{itemize}

\subsection{\textbf{Technologies}}
\begin{figure}[htb!]
    \centering
    \includegraphics[width=0.7\linewidth]{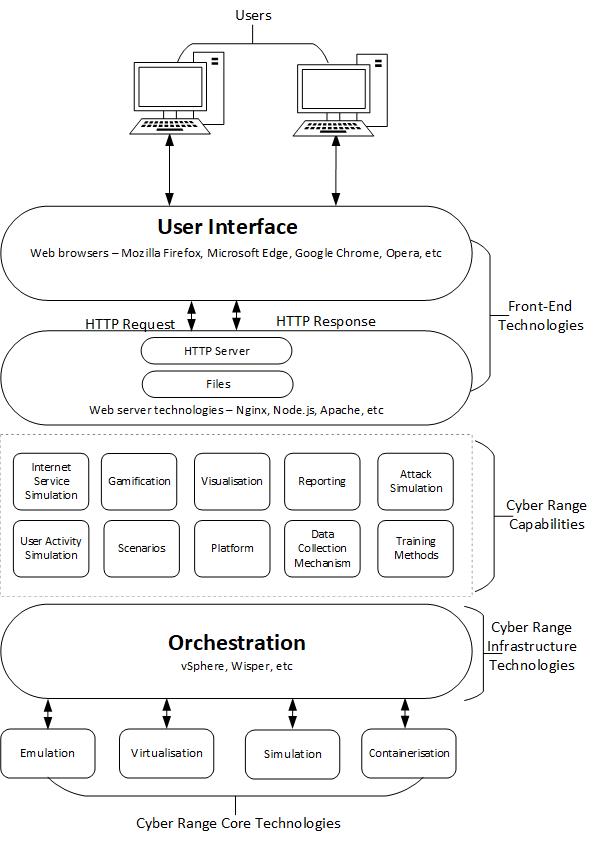}
    \caption{\textbf{Architectural Design of a Cyber Range}}
    \label{fig:CRArch}
\end{figure}

\begin{figure}[htb!]
\includegraphics[width=0.7\linewidth]{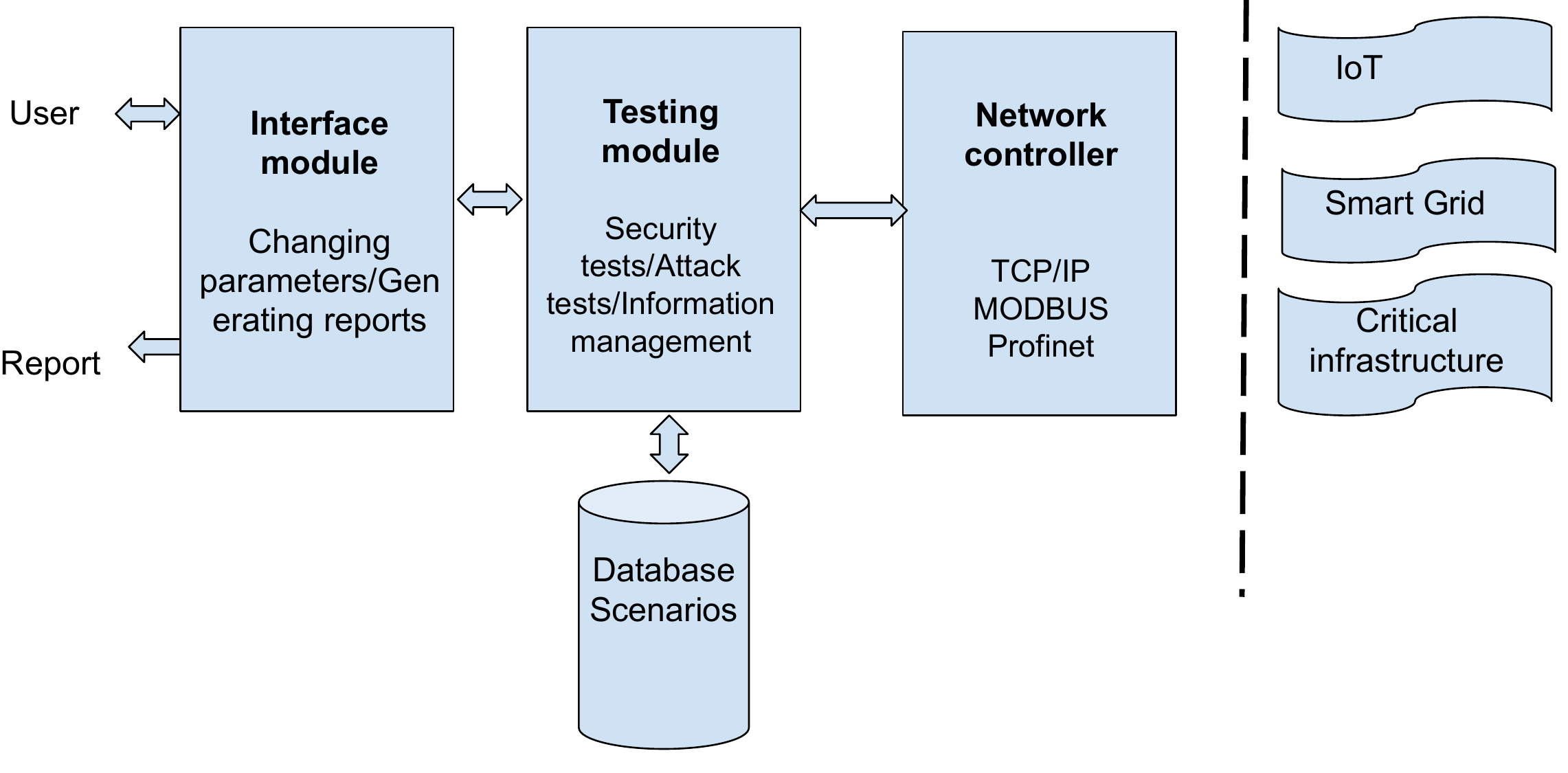}
\centering
\caption{\textbf{Architectural Design of a Test-Bed}}
  \label{fig:Testbed}
\end{figure}

 Figure~\ref{fig:CRArch} and Figure~\ref{fig:Testbed} summarise CR Core technologies segmented as virtualisation, simulation, containerisation, and physical hardware; some CRs provide a combination of these technologies such as virtualisation with physical hardware. TB implementations target the training of cyber situational awareness for domain experts in the areas of control and information technology networks. The platforms also enable training in operational technologies with few employing simulation and emulation but the bulk are based on physical hardware. Table~\ref{tab:CRTechTable} presents an overview of CRs/TBs technologies used based on the available literature with focus on the selected application areas.

\subsubsection{\textbf{Core Technologies}}
\begin{figure}[t]
    \centering
    \includegraphics[width=0.7\linewidth]{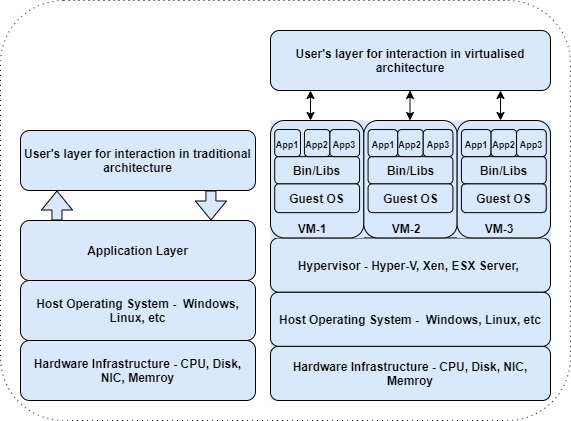}
    \caption{\textbf{Hybrid Computing Stacks}}
    \label{fig:virtualisation}
\end{figure}

The modelling of certain infrastructures underpinning a particular application require the use of a combination of methods, in effect a hybrid implementation as shown in Figure~\ref{fig:virtualisation}, where the combination of Virtualisation with Physical Hardware technologies is presented. These combinations enhance the capabilities of CR - by allowing operational and information technologies to be part of a scenario - embody features of both CR and TB.

\begin{figure}[hbt!]
    \centering
    \includegraphics[width=0.5\linewidth]{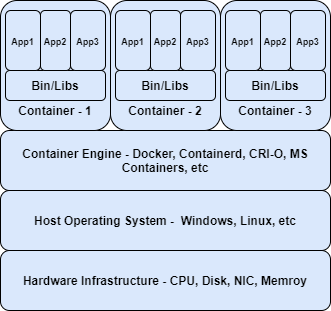}
    \caption{\textbf{Containerisation Technology}}
    \label{fig:containerisation}
\end{figure}

\begin{figure}[hbt!]
    \centering
    \includegraphics[width=0.5\linewidth]{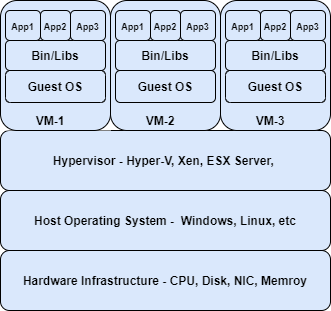}
    \caption{\textbf{Virtualisation Technology}}
    \label{fig:virtualisationTech}
\end{figure}

Containerisation such as Docker is summarised in Figure~\ref{fig:containerisation}. Containerisation is a light-weight approach to virtualisation, a uniform structure in which any application can be containerised (stored), transported, and deployed (run). Hardware virtualisation, on the other hand, implies Virtual Machine~(VM) deployment i.e., a layer between the hardware and the host operating system, managed by a hypervisor as shown in Figure~\ref{fig:virtualisationTech}. The use of containers is more scaleable compared to VMs, but the latter provide a more flexible and secure system. Their application depends largely on need but there is the possibility of VMs and Container technologies merging into a form of cloud portability.

Emulation replicates the operations within the target infrastructure through a mirror system, while simulation replicates the behaviour of the target system through a model; thus, simulation is preferred in virtual training applications.

\newpage
\subsubsection{\textbf{Infrastructure Technologies}}
Technologies that establish, manage, and control CRs are located between the core and front-end layers (Figure~\ref{fig:CRArch}). 
Their selection vary widely based on CR developers' preference and the goal application of the CR. The range of technologies are readily available, example being virtualisation management solutions such as vSphere and Wisper.

A number of CR implementations~\cite{ranka2011national}~\cite{ferguson2014national} utilise a combination of physical servers with virtual solutions. In these cases, the physical server has direct and exclusive access to the physical hardware, and the virtualisation has virtual hardware emulated by the hypervisor, which in turn controls all access to the underlying physical hardware. Virtualisation acts as a layer in between the hardware and the host operating system. Two types of hypervisor are in routine use, referred to as Type 1 and Type 2. Type 1 hypervisor runs directly on the host machine's physical hardware, while Type 2 - more commonly known as a hosted hypervisor - is installed on top of an existing operating system. A hypervisor employs four main virtual resources; vCPU, vMemory, vNetwork (vSwitch), and vDisk. 

SCADA-based TBs employ Human Machine Interfaces~(HMI) server software, software-based Relay Terminal Units~(RTUs) and Relay Programmers, as a consequence of the need to reproduce an exact model of the inter-dependencies between components. Accuracy of the model is essential in the evaluation of the effectiveness of cyber-attacks and their corresponding countermeasures~\cite{hahn2010development}~\cite{10.1007/978-3-030-12786-2_1}.

Since many CR articles do not reveal the underlying infrastructural technology in use, e.g. vSphere, Wisper in their design and implementation, the use of `Available' in the Table~\ref{tab:CRTechTable} indicates an infrastructure technology in use that cannot be specified, while `Not Available' indicates that no information on the infrastructure technology was reported. Both are included for completeness.\\

\subsubsection{\textbf{Front-End Technologies}}
\begin{figure}[hbt!]
    \centering
    \includegraphics[width=6cm]{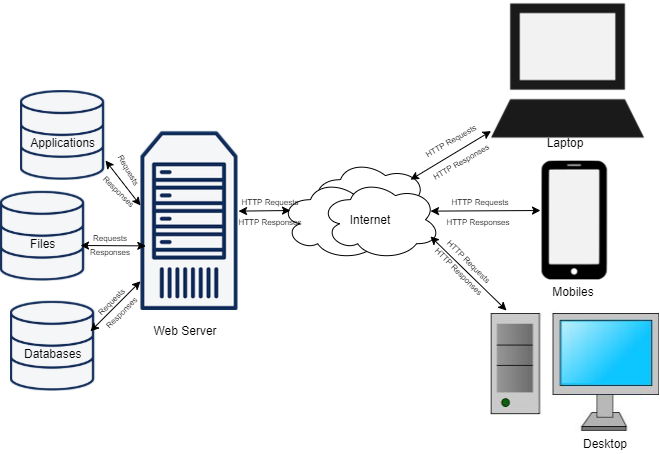}
    \caption{\textbf{A Simple Web Server Architecture}}
    \label{fig:webserver}
\end{figure}

The bridge between end user and the CR - Core and Infrastructure - is the Front-End; the Core, infrastructure and user type determine the features of the Front-End. The basic elements of a web server as shown in Figure~\ref{fig:webserver} represent the Hardware and Software components, the former is the physical server used by the hosting providers and the latter comprises an operating system and Hyper-Text Transfer Protocol~(HTTP) server databases and scripting languages that enhance the capabilities of the web server. A Web server such as Apache or Nginx is deployed at the back-end coupled with a Content Management System~(CMS) compiling results from scripting languages, databases, and HTML files to generate content for to the user. Web technologies provide the front-end interface as shown in Figure~\ref{fig:CRArch} such as HTML5-based console simulators. 

TBs, on the other hand, rarely use front-end technologies as the environments being modelled are predominately Operational and Information Technology~(OT/IT) systems on Human Machine Interface~(HMI) servers, Historians, Software-based Remote Terminal Units~(RTUs) and Relay Programmers~\cite{hahn2010development}. Table~\ref{tab:CRTechTable} provides an overview of the available literature on CR Core, Infrastructure and Front-end technologies.

\begin{table*} [ht!]
\begin{center}
\caption{\textbf{CR and TB Technologies}}
\begin{tabular}{|l|c|c|c|c|c|}
\hline
Classification&	Ref.& Year&	Core Technology&	Infrastructure Technology&	Front-End Technology\\
\hline
Cyber-SHIP&\cite{tam2019cyber}&2019&Live Scenario&Not Available&Not Available\\
\hline
ICSRange&\cite{giuliano2019icsrange}&2019&Simulation&Not Available&Not Available\\
\hline
Clusus&\cite{hildebrand2019clusus}&2019&Simulation&Available&Available\\ 
\hline
Testbed@TWISC &\cite{tsai2018testbed}&2018&Emulation&Available&Available\\
\hline
CYRAN&\cite{hallaq2018cyran}&2018& Hybrid&Available&Available\\
\hline
INSALATA &\cite{herold2017achieving}&2017&Emulation&Available&Available\\
\hline
Virginia CR	&\cite{radziwill2017virginia}&2017& Simulation& Available & Available\\
\hline
CyberVan &\cite{chadha2016cybervan}&2016&Simulation&Available&Available\\
\hline
SoftGrid &\cite{gunathilaka2016softgrid}&2016&Emulation	&Available&Available\\
\hline
SCADA-SST &\cite{ghaleb2016scada}&2016&Simulation	&Available&Available\\
\hline
KYPO&\cite{vceleda2015kypo}\cite{vykopal2017kypo}&2015&Simulation&Available&Available\\
\hline
CRATE&	\cite{sommestad2015experimentation}&2015&Emulation& Available & Available\\
\hline
DoD CR&\cite{ferguson2014national}&2014&Simulation& Available & Available\\
\hline
SCADAVT-A&\cite{almalawi2013scadavt}&2013&Live Scenario& Available & Available\\
\hline
StealthNet&\cite{varshney2011live}&2011&Simulation&Available&Available\\
\hline
NCR (DARPA)&\cite{ranka2011national}\cite{ferguson2014national}&2011&	Emulation& Available & Available \\
\hline
PowerCyber	&\cite{hahn2010development} &2010&Simulation&Available&Available\\
\hline
Reassure	&\cite{thomas2009mandatory}&2009&Simulation&Available&Available\\
\hline
ATC	&\cite{brueckner2008automated}&2008&Live Scenario&Not Available&Not Available\\
\hline
CAAJED&\cite{mudge2008cyber}&2008&Simulation& Available & Available\\
\hline
DETER &\cite{benzel2006experience}&2006&Emulation&Available&Available\\
\hline
RINSE&\cite{liljenstam2005rinse}&2005&Simulation&Not Available&Not Available\\
\hline
ViSe&\cite{richmond2005vise}&2005&Emulation&Available&Available\\
\hline
NetENGINE&\cite{brown2003simulation}&2003&Simulation&Available&Available\\
\hline
LARIAT &\cite{rossey2002lariat}&2002&Hybrid	&Available&Available\\
\hline

\end{tabular}
\label{tab:CRTechTable}
\end{center}
\end{table*}

\newpage
\section{\textbf{Scenarios and Applications}}
Different scenarios and application areas of CRs and TBs technologies will be the foc of this section.
\label{Section: ScenariosandApplications}
\subsection{\textbf{CRs and TBs Scenarios}}
\begin{figure}[hbt!]
    \centering
    \includegraphics[width=0.6\linewidth]{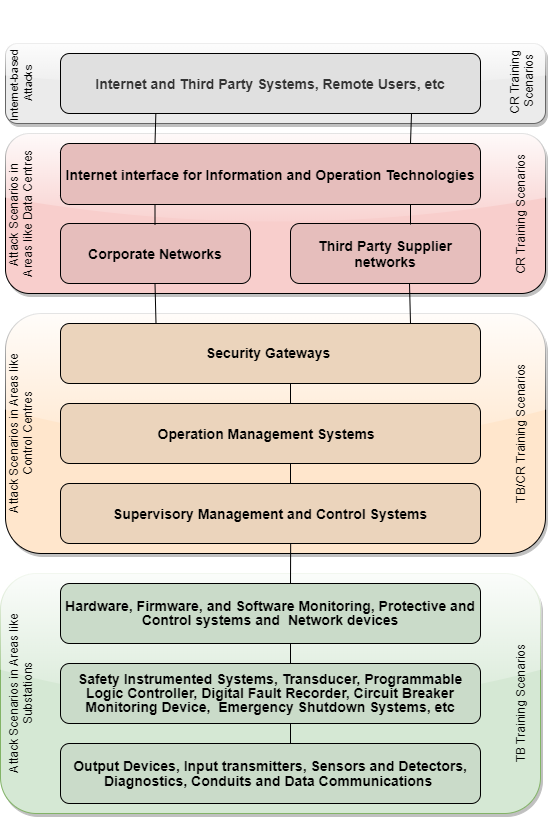}
    \caption{\textbf{Attack scenarios types}}
    \label{fig:Scenarios}
\end{figure}

Scenarios are simulated or emulated networks comprising traffic as well as potential threats in the network layer (PAN, LAN, MAN, WAN), software and hardware implemented through virtual machines (VMs), Containers or Sandboxes. In a bid to comprehensively represent target networks, the scenario can also feature other system peripherals and appliances. The simulated network environment is injected with traffic representative of user activities e.g. web surfing, email, and other server communications and real-life attack scenarios such as in Control or Data centres (Figure~\ref{fig:Scenarios}) are deployed. A predefined attack scenario library as well as custom-built scenarios are integral to the platform.

Yamin~\textit{et al.}~\cite{yamin2019cyber} state that scenarios consist of Purpose, Environment, Storyline, Type, Domain and Tools, features to appropriately classifying a scenario aligned with the objectives of the exercise/training. The major differences between TB and CR scenarios are in the attack scenarios being simulated or emulated (Figure~\ref{fig:Scenarios}). TBs predominately simulate attacks in critical infrastructures such as energy sub-stations e.g. re-configuring a relay systems/devices Denial of Service (DoS), modifying/disrupting valid alarms, producing fake alarms, sending incorrect commands to the relay, manipulating readings from a relay, and injecting incorrect data to historian~\cite{hahn2010development}, whilst CRs most often simulate  multi-connected network such as Control Centres, Data Centres, and Internet-enabled IT/OT system attacks e.g. SQL Injection, Apache Shutdown, Web Defacement, Trojan Data Leakage, Java Network Monitoring System~(NMS) Kill, Database~(DB) Dump via File Transfer Protocol, Ransomware, DDoS, Synchronise~(SYN) Flood, SCADA Human Machine Interface~(HMI).

Scenarios depend largely on the application and the architecture of the network and adapt to the training goals. The relationship between the goals of the training and the optimum scenario remains fundamental in the assessment of the positive value of CR or TB. 

Table~\ref{tab:Scenarios classification} presents a list of attacks and their associated settings in the last five years. Attacks are classified by scenario complexity and type; `Low' for scenarios with at least one attack test; `Medium' for scenarios with two classical attacks; and `High' for sophisticated or more than two attacks.

\subsubsection{\textbf{Scenario Design, Validation and Deployment}}
\begin{itemize} 
    \item \textbf{Design}: The definition of functional and non-functional as well as user and team-related requirements are essential pre-requisites in the design of a CR scenario. While the functional requirements pertain to the services the system provides, non-functional requirements describe how the system reacts to inputs and its dynamic responses. The team-related requirements are the tools and resources inherent within the exercise for use by teams.~\cite{marrocco2018design}. Furthermore, \textit{ab initio} a set of attack trees based on an understanding of how an attacker can gain access to the domain under study is imperative to an effective attack scenario design. Thus a comprehensive vulnerability assessment must be established, and coupled with the impact scenarios, are combined produce a set of attack trees, the foundation for establishing a representative real-life breach condition and in turn enabling an evaluation of the optimum countermeasures to arrest the attack~\cite{hahn2010development}\cite{ten2007vulnerability}. \\
    
    \item \textbf{Validation}: Russo~\textit{et al.}~\cite{russo2018scenario} report on a framework for automating model validation of scenarios through a Scenario Definition Language~(SDL) on the OASIS Topology and Orchestration Specification for Cloud Application~(TOSCA)~\cite{binz2014tosca}. SDL/TOSCA based implementations automate the validation of the scenario against specified design errors, such as incorrect hardware/software bindings. The approach translates a SDL design into a Data Log specification, before verifying if the specification satisfies the goals of the scenario. A design modification is triggered whenever the validation fails, otherwise the scenario is automatically deployed. While developed for CR applications, the solution is also applicable to TBs but is dependent on the domain of study, most relevant in attack scenarios in targeting Control Centres (Figure~\ref{fig:Scenarios}). \\
   
    \item \textbf{Deployment}: A number of other approaches to activating scenarios have been reported. CRACK~\cite{russo2020building} are a SDL/TOSCA scenario definition, design and deployment languages and Automated Deployment of Laboratory Environments Systems (ADLES)~\cite{de2018adles}, an open source specification language and associated deployment tool, achieve the same goals. ADLES provides an instructor a tool-set to design, specify, and semi-automatically deploy the training scenario together with tutorials as well as competitions. Furthermore, efficient sharing of classes  together with the associated computing environment are provisioned to participants. The ADLES deployment begins with the verification and fixing of Master instances by converting them into templates followed by the use of these instances to clone services, create virtual networks and folders. The full exercise scenario on the specified virtualisation platform is then deployed. While these implementations are current state-of-art deployments, it is important to acknowledge that within the foreseeable future, the effectiveness of these tools will be diluted as the sophistication and complexity of cyber-attacks evolve powered through AI-based and Bio-Inspired attack strategies, motivating the need to migrate to Real-Time Auto-configurable systems. \\

\end{itemize}

\subsection {\textbf{CR and TB Applications}}
Figure~\ref{fig:Scenarios} illustrates a clear trend in the convergence of CRs/TBs cyber-awareness training. While it is acknowledged that CRs cover a broader applications than TBs in the recent past, a number of domains where CRs are in particular use is becoming more evident, such as in industries for commercial purposes, education and research for academic purposes, military, defence and intelligence and in the defence of critical national infrastructure. TBs, although in use within these domains, are applied more extensively in Smart Grids and IoT architecture due to the embedded nature of HMI, Historian, RTUs, Relays implementations which better define the type of attack scenarios witnessed in these specific domains.

\begin{figure}[hbt!]
	\centering
		\includegraphics[scale=0.5]{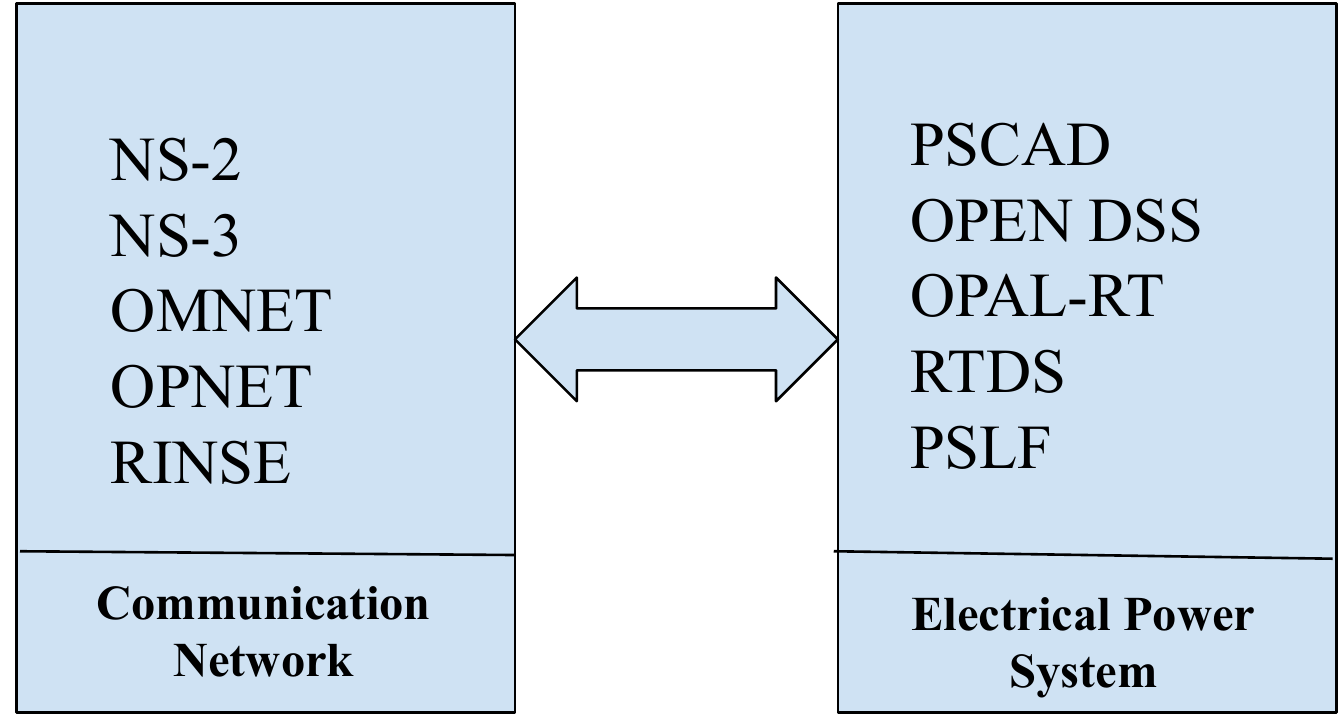}
	\caption{\textbf{Co-Simulation Test-bed}}
	\label{fig:smartG}
\end{figure}

\begin{itemize}
    \item \textbf{Industrial and Commercial}: IBM X-Force Command Centre (\cite{priyadarshini2018features}) is the first commercial malware simulator that tests for the security of systems. At the heart of the simulator is a mobile Command Cyber Tactical Operations Center~(C-TOC) that provides cyber-range and watch floor services. The C-TOC can be configured both as an immersive training CR, a platform for Red teaming and capture-the-flag competitions, as well as a watch floor for special security events. The Ixia Breaking Point, advertised as providing CR capabilities~\cite{davis2013survey},is also a commercially available. The single rack-mountable appliance provides traffic generation and a `Strike Pack' of network security and malware attacks. Exata is yet another commercially available simulation-based CR. A number of emulation-based CR are currently on offer, a good example being the ATC~\cite{brueckner2008automated}.
    \\\item \textbf{Education and Research}: Cohen~\cite{cohen1999simulating} presents the development of SECUSIM~\cite{park2001secusim}, a highly customisable system with integrated Graphic User Interface~(GUI) capabilities, the first example of the education and research community creating a training platform for simulating the impact of attacks on computer networks~\cite{davis2013survey}. The University of Illinois has developed the Real Time Immersive Network Simulation Environment~(RINSE) in 2006, also primarily for training~\cite{liljenstam2005rinse}. Other implementations in academia for modelling computer networks and intrusion detection systems~(IDSs) attacks include the Virginia CR~\cite{radziwill2017virginia}, Emulab~\cite{siaterlis2012use}, Virtualised CR~\cite{mayo2009approaches}, ARENA~\cite{kuhl2007cyber}. NetENGINE~\cite{brown2003simulation} has been designed for training on the strategies to combat cyber-attacks in large IP networks comprising a Virtual Cyber-Security Testing Capability~(VCSTS) for the automated testing of new devices to assess its security robustness before deployment~\cite{pederson2008virtual}.\\
    
    \item \textbf{Military, Defence and Intelligence}: Davis and Magrath~\cite{davis2013survey} assert that the USA Air Force (USAF) used CR around 2002, an element of the Simulator Training Exercise Network~(SIMTEX) referred to as the Black Demon. The first reported CR was the Defence Advanced Research Projects Agency~(DARPA)’s National Cyber Range~(NCR) representing the foundation in the training of their military, defence and intelligence agencies on cyber warfare initiated by the United State military in 2009 as a consequence of the US Department of Defence classified military computer networking infrastructure being significantly compromised in 2008~\cite{lynn2010defending}. Although NCR was largely a military-sponsored initiative, its use and application cut across the military, commercial, academic, and Government sectors~\cite{davis2013survey}~\cite{ranka2011national}. Fourteen (14) CR applications in Military, Defense and Intelligence have been recorded to date ranging from CRATE~\cite{sommestad2015experimentation}, DoD CR~\cite{ferguson2014national}, CAAJED~\cite{mudge2008cyber}, SAST~\cite{meitzler2009security}, StealthNet~\cite{varshney2011live}, LARIAT~\cite{rossey2002lariat} to INL~\cite{anderson2009cyber}. Their role is not only to train the security agencies of sovereign countries on counter cyber-terrorism and warfare, but also to protect the nation's critical infrastructure such as Naval, Power and Aviation. SoftGrid~\cite{gunathilaka2016softgrid} and CyberVan~\cite{chadha2016cybervan} are examples of TBs found in these application sectors.\\

    \item \textbf{Smart Grids}:
    The predominate area of application for TBs is Smart Grids owing to the reliance for the effective operation of an ever-evolving power network on an enabling communication network with information flow managing the power delivery. Consequently, the security of equipment and the critical signals that control the power system becomes essential for the safe, flexible and uninterrupted provision of the supply of energy.

    A Smart Grid Test-bed can be cast as two simulation environments (Figure~\ref{fig:smartG}), one for the power, the other for the cyber/communication network. Co-simulator segmentation is a necessity as a hacker can target operations within both networks~\cite{10.1145/3407023.3409313}. Here, TBs are classified into two categories; off-line and real-time. An off-line environment is the most prevalent approach realised, most readily, by SCADA systems~\cite{hammad2019implementation}; refer to Table~\ref{tab:OfflineAndRealTimeTestbeds} for details. OMNET++ or NS2 are invariably at the core of most cyber simulators, with the TCP/IP protocol used to communicate between simulators. Synchronisation is central to the co-ordination of operations in the two domains. Real-time TBs have been proven to facilitate efficient training outcomes~\cite{poudel2017real}.\\
    
    \item\textbf{IoT Devices}:
    In the recent past, Internet-of-Things (IoT) architectures have evolved rapidly characterised by a growing complexity of inter-connections of an ever-increasing number of nodes (`things'). The proliferation of highly connected environments translates into an enhanced spectrum of vulnerabilities/opportunities for cyber criminals. Furthermore, IoT-inspired data-driven solutions have been adopted by key industry sectors as a means to implement business transformation. Securing network infrastructures consisting for example, of medical records, financial credential information against breaches becomes even more challenging. The training of security operators in these new classes of threats is essential. IoT Test-beds that can simulate different kinds of attacks play an important role in supporting the delivery of dynamically changing training requirements. Example IoT Test-Beds have been reported in~\cite{waraga2020design},~\cite{kim2019soda} and~\cite{lee2018design}.\\

\end{itemize}

\begin{table}[hbt!]
\caption{\textbf{Offline and Real-Time Test-Beds}}
\scriptsize
	\centering
		\begin{tabular}{|c|c|}
		\hline
			 \textbf{Real-time} & \textbf{Offline}\\
			\hline
			Expensive & Less expensive\\
			\hline
			Complex & Easy implementation\\
			\hline
			Time-gain & Extended time\\
			\hline
			Integrate generators\ controllers & Cannot integrate hardware systems\\
			\hline
		\end{tabular}
	
	\label{tab:OfflineAndRealTimeTestbeds}
\end{table}

\begin{table*}[hbt!]
\caption{\textbf{Some Cyber-Attacks and their Domain in the last 5 years}}
\footnotesize
\centering
		\begin{tabular}{|p{0.5cm}|p{0.5cm}|p{2.0cm}|p{2.4cm}|p{1.2cm}|p{3.0cm}|p{4.0cm}|}
		\hline
			 \textbf{Ref} & \textbf{Year}& \textbf{Domain}& \textbf{Tool}& \textbf{Complexity}& \textbf{Attack type}& \textbf{Commentary}\\
			\hline
			\cite{wang2020distributed} & 2020& Smart Grid& OPAL-RT& Medium& test-bed cyber events& Cyber attack needed to validate\\
			\hline
			\cite{waraga2020design} & 2020& IoT& Open source platform&High& Extensive analysis/Automated tests & Time analysis needed to show the efficiency of Test-bed\\
			\hline
			\cite{hammad2019implementation} & 2019& Smart Grid (PSCADA)&OMNET& Low&DoS/FDI& Lack of testes and scenarios\\
			\hline
			\cite{de2019implementation} & 2019& IoT (SCADA)&--& Medium& DNS attack& security tests required\\
			\hline
			\cite{kumar2018secure} & 2019& IoT&QEMU emulator& Low& DDOS attack&Few attacks are tested\\
			\hline
			\cite{alves2018virtualization} & 2018&Water storage (SCADA)&--& High& Packet injection/ARP spoofing/DoS&Real-time implementation discussion missed\\
			\hline
			\cite{siracusano2018framework} & 2018& Information centric network&CONET& High&Input traffic pattern&Adding further experiences\\
			\hline
			\cite{poudel2017real} & 2017& Smart Grid& OPAL-RT& High&Access to communication link&Real-time implementation of OPF model\\
			\hline
			\cite{papadopoulos2017thorough} & 2017&IoT&FIT/IoT LAB& Low& No attack tested& lack of security tests\\
			\hline
			\cite{liu2017cyber} & 2017&Cloud&Open source& Low& No attack tested& lack of security tests\\
			\hline
			\cite{bernieri2017monitoring} & 2017&Industrial control system (SCADA)&--& Medium& Availability attack/Integrity attack& Detection tool should be implemented\\
			\hline
			\cite{lee2018design} & 2017&IoT(SCADA)&Hardware-based test bed& High& 5 kinds of attack& Test bed with IDS\\
			\hline
			\cite{flauzac2016developing} & 2016&IoT&Software-based OpenFlow switches& Low& No attacks& Software defined networking testbed\\
			\hline
			\cite{ashok2016testbed} & 2016&Smart Grid&Real Time Digital Simulator& Low&Man-in-the-Middle attack& Test bed with  Attack Resilient Control algorithm\\
			\hline
			\cite{ghaleb2016scada} & 2016&SCADA&C++& Low&Denial of Service attack& Test-bed based on  SCADA simulation environment (SCADA-SST) \\
			\hline
			\cite{deshmukh2016hands} & 2016&SCADA&--& Low&Man-in-the-Middle Attack& Test-bed using CPS topology \\
			\hline
			\cite{adhikari2016wams} & 2016&Power System&Real Time Digital Simulator (RTDS)& Medium&Aurora Attack/Network Based Cyber-Attacks& WAMS cyber-physical test-bed \\
			\hline
			\cite{ashok2015experimental} & 2015&Power System&Real Time Digital Simulator (RTDS)& Medium&Measurement attack/Control  attacks& PowerCyber CPS security  Test-bed \\
			\hline
			
		\end{tabular}
	
	\label{tab:Scenarios classification}
\end{table*}

\subsection{\textbf{CR and TB Realisation}}
\begin{figure*}[hbt!]
    \centering
    \includegraphics[width=\linewidth]{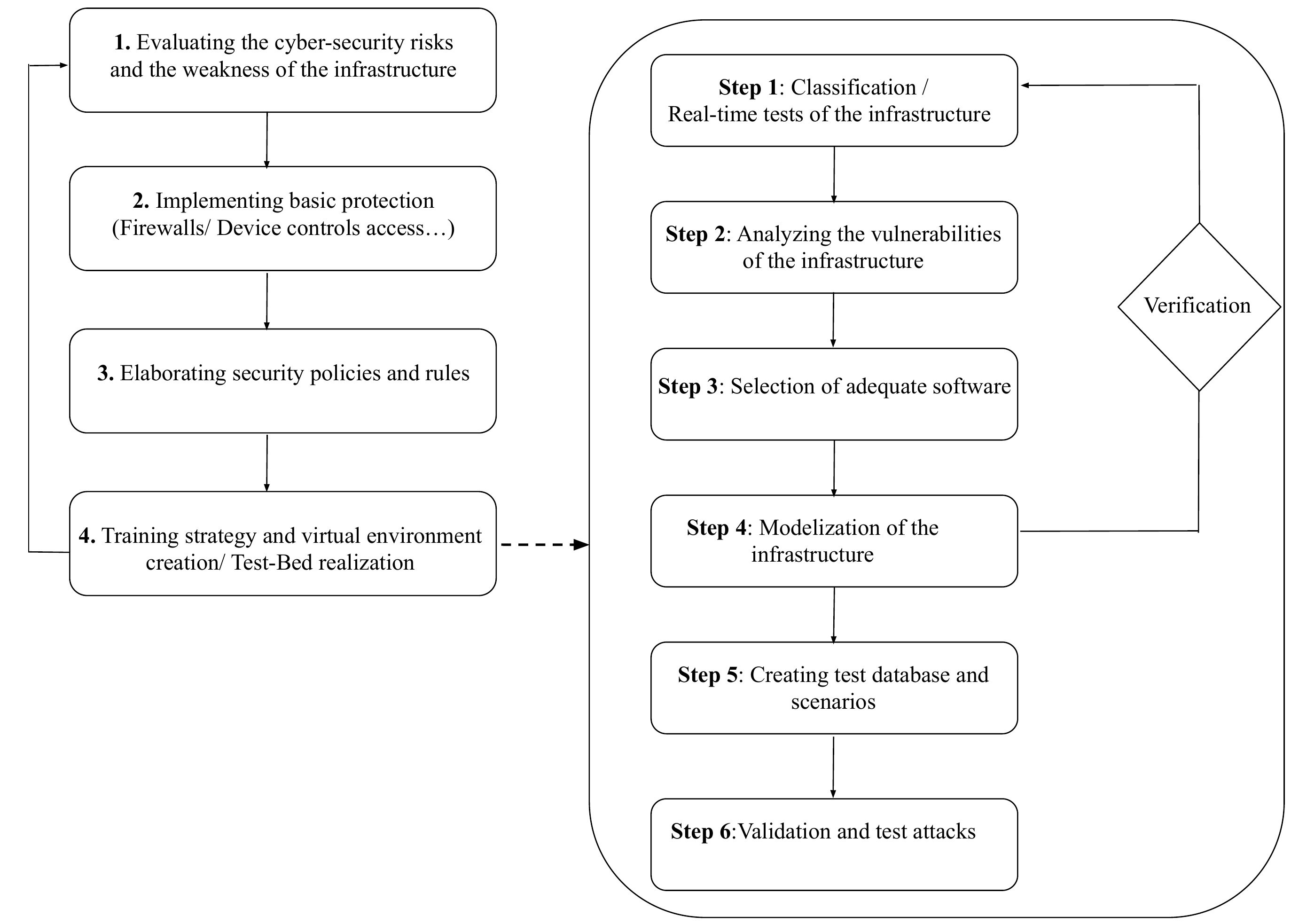}
    \caption{\textbf{Cyber-range/Test-bed flowchart}}
    \label{fig:Test-bed realization}
\end{figure*}

The flowchart in Figure~\ref{fig:Test-bed realization} describes the steps in its realisation;
\begin{enumerate}
    \item Evaluate the weaknesses in the infrastructure; the architecture of the local network, past attacks, and the current security strategy should be examined.
    \item Map the basic solutions; e.g. a firewall or control of external devices.
    \item Asses current security policies; to enhance the security level of an infrastructure.
    \item Training by simulation of attack exercises and scenarios; the definition of an appropriate virtual TB for training should adopt the following steps;
    \paragraph{Step 1}
A classification of the infrastructure is an essential step in advance of the realisation of the TB. Validation tests on the infrastructure may are required for an accurate classification.
\paragraph{Step 2}
The vulnerabilities of the infrastructure should be identified; the localisation of vulnerabilities is important in informing on the security deficiencies within the infrastructure.
\paragraph{Step 3}
Selection of the most appropriate software dependent on the application domain and the infrastructure. As an example, in the Smart Grid environment, OPAL-RT can be chosen to simulate the electric power and a discrete event network simulator to simulate the communication network.
\paragraph{Step 4}
A modular approach is adopted to describe the infrastructure, with the input/output of each module verified.
\paragraph{Step 5}
A database of different tests and scenarios is created, fundamental for the validation of the TB.
\paragraph{Step 6}
Users are trained to respond to a range of attacks and threat scenarios, with the relevant reports being extracted from the interface module of the TB.
    
\end{enumerate}

\section{\textbf{Analysis and Taxonomies}}
\label{Section: AnalysisandTaxonomies}
\begin{figure*}[ht!]
\includegraphics[width=\linewidth]{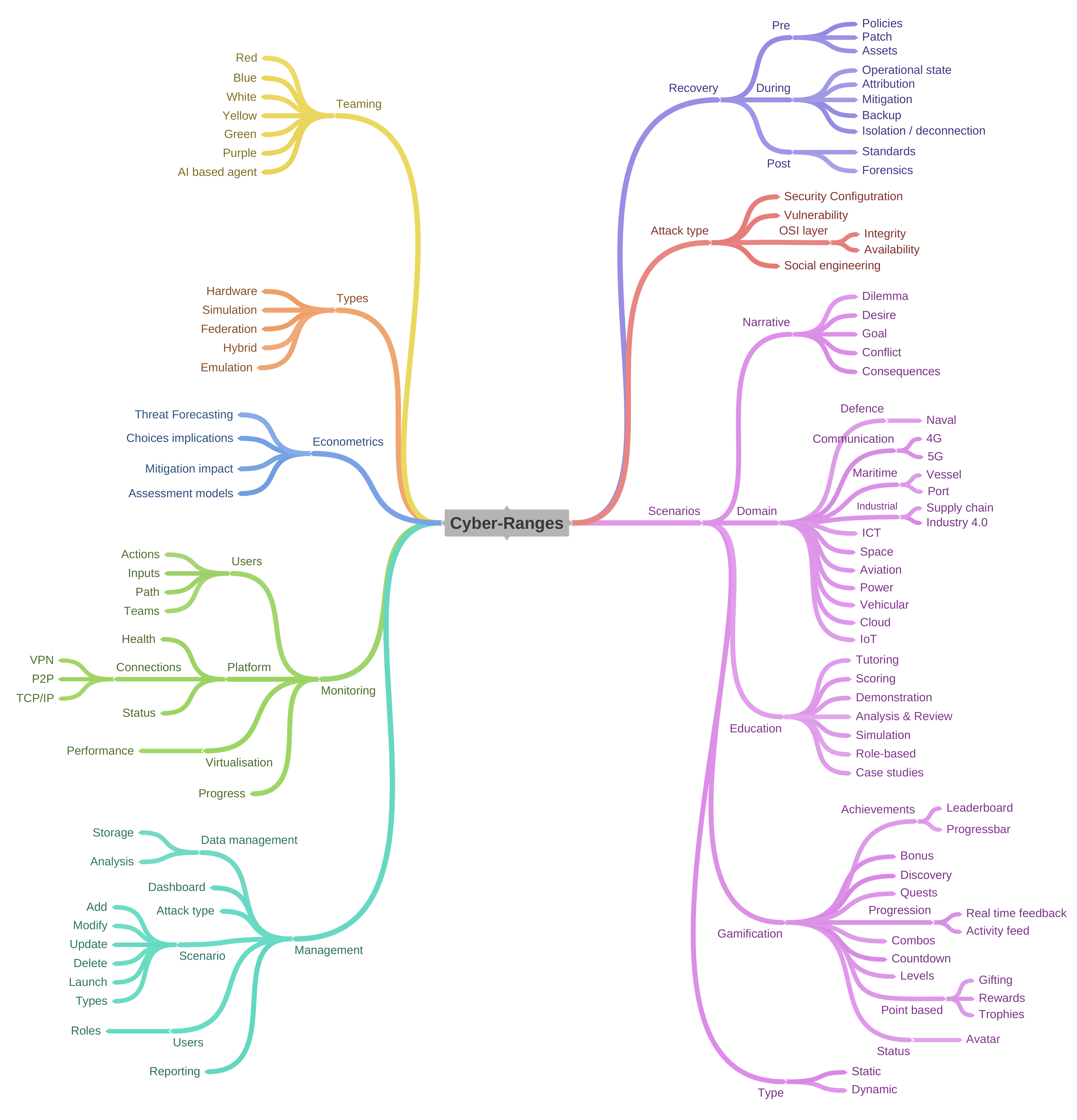}
\centering
\caption{\textbf{Cyber-Range Taxonomy}}
  \label{Fig.CR}
\end{figure*}

\begin{figure*}[htb!]
\includegraphics[width=0.85\linewidth]{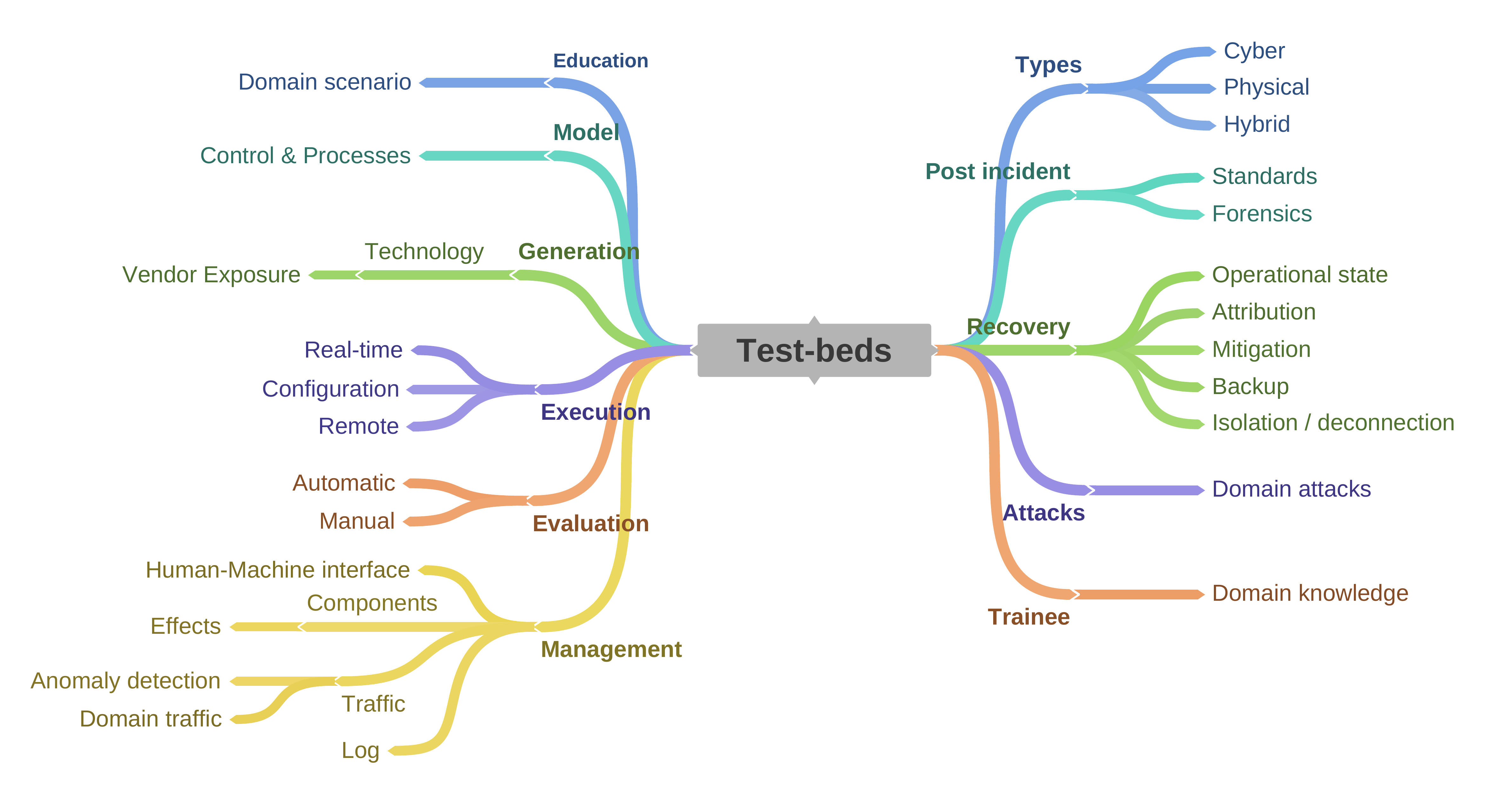}
\centering
\caption{\textbf{Test-Bed Taxonomy}}
  \label{Fig.TB}
\end{figure*}

Two taxonomies in Figures~\ref{Fig.CR} and \ref{Fig.TB} for treating CRs/TBs have been established based upon the reviewed literature. Current taxonomies encompass both CR and TB due to the close coupling between platforms, however, each offers different services governed by their implementation and training aims. The differentiation is captured in order to compile evidence demonstrating that CRs are mostly applied in IT while TB are preferred in OT environments. Moreover, CR are orientated towards end-users with a general understanding of the simulated architecture, while test-beds often require domain knowledge. The differentiation confirms the need for two separate taxonomies. 

\subsection{\textbf{Cyber-Ranges}}
The definition of the cyber-range taxonomy is informed by future developments as inferred from the reviews conducted in this paper. 

\subsubsection{\textbf{Management}}
The management layer presents a range of interfaces to various users, administering the collection, storage and analysis of the data describing scenarios and user-interactions. Information is presented to users through a dashboard along with the available scenarios and attack types per scenario. The layer also administers users and their roles as well as being responsible for reporting. 
\subsubsection{\textbf{Monitoring}} The component monitors users on the platform, capturing progress and assessing performance throughout the different scenarios as well as being responsible for connections of remote users to the platform, their actions, inputs paths selection and team formations.  This component also validates the health of the platform and the various services and scenarios provisioned. 
\subsubsection{\textbf{Econometrics}}
Understanding the impact of the actions taken by an user is essential, especially to estimate the level of situational awareness. The component executes an evaluation of the economic impact of actions taken by users within the various scenarios. 
\subsubsection{\textbf{Types}}
Hardware based CRs allow training on operational technologies such as programmable logic controllers. Simulation/Emulation based CRs allow an infrastructure to be replicated, are scalable and cost effective, however, it is often challenging to replicate architecture accurately due to software limitations. A federated approach may be adopted where multiple CRs are clustered, each CR dedicated to simulating a single environment e.g. Large Enterprise Network and a Power Network and creating scenarios that span across all CRs. The hybrid solution, while similar, often depicts CRs composed of Hardware and Software solutions to provide both scalability, and affordability.
\subsubsection{\textbf{Teaming}}
Teams are at the heart of managing cyber protection services for organisations and consequently CRs are required to provide the appropriate environments for appropriate training. The \textit{Yellow} team comprises application developers and software architects managing the CR. The \textit{Green} team focuses on enhancing the security provision, the automation of tasks and ensure that the code is of the highest quality. The \textit{Orange} team facilitates the education and is responsible of the creation and development of scenarios. The \textit{Blue} team focuses on developing defensive actions, to protect the network and define the most effective countermeasure to arrest the breach. The \textit{Red} team adopts an offensive stance, often competing against the Blue team. Finally, the \textit{Purple} team is composed of users with both Blue and Red team skills, with knowledge of both defensive and offensive tactics. 
\subsubsection{\textbf{Recovery}} The recovery component ensures that all policies and patches remain up to date. The component maintains the operational state of the CR during an exercise, executes regular back-ups and restricts cyber-attacks spilling from the CR. The function is central for digital forensic purposes post incident/cyber-attack. 
\subsubsection{\textbf{Attack Types}} The component encompasses descriptions of the different attacks including the security configurations for the vulnerabilities within scenarios. A database of the vulnerabilities, as well as a high/low level description of each mapped against the OSI model is established.
\subsubsection{\textbf{Scenarios}} The scenario component is subdivided in five sub-components focusing on I) the Narrative - it is essential for a scenario to have a target goal as well as the consequences of any action. A desire, dilemma and conflicts can also be added to enrich the learning environment.  II) the Domain defines the context in which the scenario is currently being simulated. III) the Education supports users to navigate and learn the skills necessary to complete the scenario through tutoring, scoring, demonstration, analysis and review of actions with the user in a role base fashion or through a specific case study. IV) Gamification is used to embed game mechanics to drive and maintain the level of user engagement e.g. encourage users to engage with the platform and/or to perform a specific task by enticing with a lure aligned to user behaviour/preferences. V) the type of scenario can be either static with a single goal or dynamic evolving with each action of the user. 
\subsection{\textbf{Test-Beds}}
In line with the CR taxonomy discussed earlier, the focus of the proposed TB taxonomy is also informed by future developments/technologies.

\subsubsection{\textbf{Education}}
The Education component is used to explore new security scenarios, most often utilised by the evaluation team to develop and confirm the scenario for the optimum learning outcomes best students. Such exercises may include the evaluation of the formative assessment and ease of implementation.
\subsubsection{\textbf{Model}}
The Modelling component provides control as well as directs process on the innovation cycle. A model of the innovation is created and processed in a controlled environment satisfying a set of constraints.
\subsubsection{\textbf{Generation}}
The Generation component provisions comprehensive information on the underlying technology and vendors, inputs that inform the features of the innovation and its deployment.
\subsubsection{\textbf{Execution}}
Real-time, configuration-based remote creation of innovation provide insights into its impact on the behaviours of the system being modelled or tested. Essential to the test of the resilience of the targeted system is an evaluation of the  behaviour at different execution scenarios, optimally executed within a controlled environment as that provided by a test-bed.
\subsubsection{\textbf{Evaluation}}
Evaluation of model within a TB can be done manually or automatically. The former is executed with human intervention, the latter harnesses an algorithm established with considerations of the key variables of the system.
\subsubsection{\textbf{Management}}
The Management components like CRs, present a number of interfaces as a function of the type of users. The services ranging from managing human-machine interface between the user and the TB helping to mitigate the limitations of these interactions, to managing the traffic for anomaly detection and representative domain traffic. The module also provides log statistics of user activities, generates reports and feedback. For example, SCADA-based TBs employ Human Machine Interfaces server software, software-based Relay Terminal Units and Relay Programmers. An accurate model of the inter-dependencies between the energy and cyber components is essential to the evaluation of the impact of cyber-attacks and in informing on the most effective countermeasure.
\subsubsection{\textbf{Types}}
Cyber-based TBs test innovation in an Internet-enabled environment; stand-alone physical TBs operate within an controlled environment, isolated from an operational network. The hybrid TB solution is a combination of Cyber and Physical TBs, comprising hardware and software in a networked as well as an isolated environment to provide training in OT, scalability, and affordability.
\subsubsection{\textbf{Post-Incident}}
The component ensures the integrity of the post incident procedures, the basis for an investigation of the performance of an innovation as well as confirming the validity of the process used in testing an attack or a failure of an innovation.  Standard and Forensics are two types of Post-Incidence investigation, the former used to provide a detailed review that helps to understand each phase of an incident, from start to finish. In a situation awareness review, such components are one step in the incident response process that requires a cross-functional participation from all individuals to determine the root cause and full scope of the attack. Forensic, on the other hand, enables a scientifically derived and proven method to collect, validate, identify, analyse and interpret evidence derived from digital sources. An evidence-based review that characterises an incident from start to finish is generated.
\subsubsection{\textbf{Recovery}}
Recovery ensures that all policies are up to date, that the operational state is maintained and that regular back-ups are being carried out. The component is also of use for digital forensic purposes after an incident, helping to mitigate further failure or attack. Furthermore, in the process of surfacing the root causes of failures, it helps in isolating and disconnecting the system under investigation.
\subsubsection{\textbf{Attacks}}
The Attack component encompasses descriptions of potential attacks including the security configurations for the vulnerabilities within scenarios. A  database of the vulnerabilities is created together with a high/low level description of each vulnerability mapped against the OSI model.
\subsubsection{\textbf{Trainee}}
The Trainee component contains specific domain knowledge required for  and records the progress of each trainee with regard to specific modules and performance measures. A report is usually displayed in the trainee dashboard.


\section{\textbf{Training Methods}}
\label{Section: TrainingMethods}
The spine of the training is founded on strategies informed by educational methodologies and is most often segmented into two classes. The first is centred on the relationship between coach and trainee using classical training methods characterised by the use of a number of support tools such as online courses, certification, training, and presentation. The second method relies more heavily on new elements such as gamification and video-assisted techniques.\\

\textbf{Classical Training:}
The fundamental goal is to train trainees to acquire new skills. In the cyber-security context, the theoretical background and knowledge of security terminologies is considered the minimum level of achievement. In general, the information flow between a coach and a trainee is one-way. For instance, online courses and presentations which, for example, describe the architecture of an infrastructure is such a case, the trainee being a passive information recipient. 

Classical training methods adopt a three-prong approach to learning ranging from getting acquainted with facts, followed by logical tools for the organisation of facts, culminating in the ability to critically analyse and draw conclusions~\cite{bauer1999classical}. The methodology inculcates the ability to comprehend and take timely and appropriate actions in dealing with cyber-related malicious activities both at the technical and operator level. The resultant knowledge on the successes and failures inherent in cyber defence scenarios, is central to a comprehensive cyber situation awareness training program in both the public and private sectors~\cite{brynielsson2016cyber}.\\

\textbf{CR and TB Training Methods}
Simulation environments implemented through CRs are one of principle routes to establishing realistic scenarios of target systems, facilitating training through a rich illustration of real-life security incidents and threats dynamics, thereby preparing and equipping operators in the selection of the most appropriate responses. The predominant training role of TBs is to emulate the impact of a range of attack scenarios and test the strategies to arrest such attacks. The trainee is able to modify the parameters of attacks, test the effectiveness of responses and extract an analysis from the output reports. The result is an assessment of the security level of the infrastructure as a function of different attack scenarios. TBs are the foundation of the practical elements of the overall training.

The commonly used strands of the training scope can be classified as:

\begin{itemize}
    \item {\textbf{Gamification:}}
     Gamification has been adopted to make cyber-security training more engaging and motivating~\cite{boopathi2015learning}. The principle is to enhance exercises through a compelling experience utilising graphics and play. The aim is to enrich the challenge, engagement, as well as motivating the trainees owing to increased levels of interaction. The concept of `Attacker-centric Gamification' was introduced by Adams and Makramalla in~\cite{adams2015cybersecurity} with the goal empowering trainees to assume the roles of attacker combining gamification with entrepreneurial perspectives with an emphasis on surfacing their abilities, skills, knowledge, motivation, and resources~\cite{10.1007/978-3-030-22351-9_7}.\\
    
    \item{\textbf{Mock Attack Training:}} The training method, developed by Sadeh~\textit{et al.}~\cite{sadeh2017mock}, embodies an approach that senses user actions which expose that user's infrastructure to cyber threats. The action could be as a result of a mock attack delivered to the user through a messaging service from any device, a wireless communication service or a fake malware application. The system selects the most appropriate training from a list of available training routines based on the users' reaction to the message in so ding delivering the most targeted training.\\
    
    \item{\textbf{Role-Based Training:}} One practical training approach through CR-enabled scenarios is to assign unique roles to trainees. Such roles, for example emulating or taking the place of a hacker in a real life situation, cyber offensive operator, cyber defender, or training instructor~\cite{toth2013role} can be dynamic depending on the exercise, defined or selected using databases that contain predefined roles. Furthermore, customisation to better emulate real-life enterprise is also possible.\\
    
    \item{\textbf{Exercises:}} Competitions such as~\cite{glumich2011defex},~\cite{conklin2005use},~\cite{dodge2004organized},~\cite{augustine2006cyber},~\cite{mattson2007cyber}, are aimed at developing problem solving techniques, proficiency, teamwork and cyber defense skills by providing the participants with sets of hands-on cyber-security exercises in real-world scenarios to the participants. `Catch-The-Flag' is an example, a distributed, wide-area security hacking competition involving multiple teams. `Cyber Defense Exercise~(CDX)' is another form of such an exercise in a larger setting where an inter-agency academy of an institution competes in the design, implementation, management and defend a network of computers. CDX is established by setting objectives, selecting an approach, defining a network topology, creating a scenario, stipulating the rules and choosing the right metrics with which to determine the lessons learnt~\cite{patriciu2009guide}.  \\
\end{itemize}

\section{\textbf{Threat Dynamics and Analyses}}
\label{Section: ThreatDynamicsandAnalyses}
\subsection{\textbf{Threat Dynamics}}
\subsubsection{\textbf{AI-based Attacks}}
The recent advances in Artificial Intelligence~(AI) has been embraced by cyber-criminals to automate attack processes~\cite{kaloudi2020ai},~\cite{brundage2018malicious}, taking advantage of technologically enhanced learning and automation capabilities offered by deep and reinforcement learning. The trend has necessitated the pressing need to develop appropriate training methods, scenarios and technologies in response.

Kaloudi and Li~\cite{kaloudi2020ai} reported a list of existing AI-enhanced cyber-attacks; (1)~Next Generational Malware such as DeepLocker~\cite{kirat2018deeplocker} and Smart Malware~\cite{cohen1999simulating}. (2)~Voice Synthesis such as Stealthy Spyware~\cite{zhang2018using}. (3)~Password-based Attacks such as Next-generation password brute-force attack~\cite{trieu2018artificial} and PassGAN~\cite{hitaj2019passgan}. (4)~Social Bots such as: SNAP\textunderscore~R~\cite{seymour2016weaponizing}, DeepPhish~\cite{bahnsen2018deepphish} and Fake reviews attack~\cite{yao2017automated}. (5)~Adversarial Training such as MalGAN~\cite{hu2017generating}, DeepDGA~\cite{anderson2016deepdga} and DeepHack. The majority of these attacks targeted interconnected and software dependent new generational embedded systems known as Smart Cyber Physical Systems such as smart traffic management systems, smart healthcare systems, smart grids, smart buildings, autonomous automotive systems, autonomous ships, robots, smart homes and intelligent transport systems.

\subsubsection{\textbf{Bio-Inspired Attacks}}
The Backtracking Search Optimisation Algorithm~(BSA) and Particle Swarm Optimisation~(PSO) are two Active System Identification attacks developed by~\cite{de2017bio} using bio-inspired meta-heuristics~\cite{farah2019image} and tested in a controlled environment. The goal was to highlight the potential impacts of automated attacks, especially their degree of accuracy in damaging the Network Controlled Systems, as a stimulus to develop solutions that counter this attack class.
Chen~\textit{et al.}~\cite{chen2016decapitation} coined the term `A Bio-inspired Transmissive Attack', a scenario exemplified in Stuxnet~\cite{langner2011stuxnet},~\cite{farwell2011stuxnet},~\cite{chen2011lessons},~\cite{lindsay2013stuxnet}, best described as a stealthy breach utilising a biological epidemic model in the communication system to propagate the attack. In addition to the hidden nature of the attack, the hacker need not be conversant with the network topology to succeed. Hence, the linkage between transmissive attacks and epidemic models.

\subsection{\textbf{Threat Analyses}}
The essence of threat analysis is to determine the potential threats, weaknesses, and vulnerabilities that can be exploited to achieve malicious goals~\cite{stango2009threat}. An understanding of the possible threats and their characteristics informs on the optimum prevention, and mitigation measures. The optimum response is also governed by the existing risk mitigation policies for a specific architecture, functionality, and configuration as defined by regulating bodies. One of the challenging requirements is the metrics to be used to determine the status of the network security performance, the basis to define approaches to increase its robustness.

\subsubsection{\textbf{Intrusion Detection System}}
 A number of Intrusion Detection Systems~(IDS) have been reported in the last decade aimed at detecting and preventing the effects of threats and network attacks. According to Hindy~\textit{et al.}~\cite{hindy2018taxonomy}~for an IDS to be considered effective, the key metrics to be measured are the high detection rates, low false positive rate, transparency, safety of the overall system, memory requirements, power consumption and throughput. However, due to the diversity of attacks and the severity of their impact, a range of IDS based on deep learning, genetic algorithms, and artificial intelligence have been developed, presented in the research papers summarised in Table~\ref{tab:IDS types}.

IDS can be classified into two groups; firstly, classical IDS based on signature detection, where only known attacks are detected and considered and secondly, anomaly-based IDS, which exploit predefined packages in training and testing~\cite{9108270}. Moreover, these kinds of IDS cannot differentiate between new attack scenarios and normal traffic. Thus, a new generation of IDS based on machine learning and artificial intelligence have been proposed, such as the work of Hodo~\textit{et al.}~\cite{hodo2016threat} on the use of Artificial Neural Network~(ANN) in IoT networks to help detect Distributed Denial of Service~(DDoS) and Denial of Service~(DoS) attacks. The aim is to distinguish automatically, without human intervention, between normal and malicious packages.

\subsubsection{\textbf{Modelling-based Approach}}
The aim is to predict the behaviours of unknown attacks and to create models able to prevent threats.  the actual vulnerability and security default of the system is core in order to conceptualise such a model, such as presented in Table \ref{tab:Testbeds}. The configuration and architecture of the local network is a requirement in the development of a cyber-threat detection model.

\begin{table}[htb!]
\begin{center}
\caption{\textbf{Attack Classifications}}
\footnotesize
\begin{tabular}{|m{0.1\linewidth}|>{
\centering\arraybackslash}m{0.05\linewidth} |m{0.4\linewidth}|m{0.4\linewidth}|}
  \hline
	
  Domain &Ref & Experimental tests / Scenarios & Tools\\
	
  \hline
	\multirow{7}{*}{IoT}& \cite{siboni2016security}&Network mapping attack/Implementation of profiling module (Training and testing algorithm) & TestStad/ Machine Learning Algorithm \\ \cline{2-4}
	
	&\cite{wang2019capacity}& Discrete-time Markov Chain model (DTMC): Analysing the capacity of the block chain& Block mining algorithm and Ethereum protocol \\ \cline{2-4}
														
	&\cite{waraga2020design}& Manual test: Analysis and attacks of each device, Automated test: process testing of different IoT device &Open Source MS \\ \cline{2-4}
    
    &\cite{lee2018design}& DoS massif trafic/Transfert Data/Abnormal code/System crash &DTM by Triangle Micro Works \\ \cline{2-4}
    
    &\cite{kim2019soda}& Real-world attack scenarios: internal and external network attacks &SDN/network function virtualisation \\ \cline{2-4}
	
	&\cite{shafiq2020selection}& Anomaly intrusion/ Attacks traffic &Machine Learning Algorithm/ Feature Extraction \\ \cline{2-4}
	
	&\cite{zolanvari2018effect}& Command injection attack &Machine Learning Algorithm/ PLC programming by Ladder language \\	 \cline{2-4}
	
	&\cite{elnour2020dual}&  SWaT/WADI datasets:Normal and attack scenario &Machine Learning Algorithm \\	 \cline{2-4}
	
	&\cite{molina2018enhancing} &  Man-in-the-middle attack  &SDN /Python \\  \cline{2-4}
	
	&\cite{arockia2019testbed}& LAUP algorithm(authentication)/ key distribution test &COOJA simulator \\
	\hline
	
	\multirow{4}{*}{Smart Grid}& \cite{hammad2019implementation} &Offline co-simulation Test-bed: DoS/FDI attacks& OMNET++ \\ \cline{2-4}
	
	& \cite{poudel2017real}& Access to communication link (\cite{hahn2013cyber}) attack model& OPAL-RT \\ \cline{2-4}
	&\cite{de2019implementation}& Deep packet inspection &Software Defined Networks/OpenFMB \\ \cline{2-4}
	
	&\cite{adepu2018epic}& Power supply interruption Attack/Physical damage attack &Real world power system/Machine learning \\		 \cline{2-4}	
	&\cite{fujdiak2019communication}& MMS/GOOSE/SV implementation &IEC 61850 Protocol/Ethernet RaspberryPi 3B+ \\	 \cline{2-4}
	
	&\cite{cheng2018development}&  HIL simulation/ proof-of-concept validation & Python \\ \cline{2-4}
	
	&\cite{liu2015integrated}&  DoS/Man in the middle attacks/TCP SYN Flood Attack &DeterLab/Security Experimentation EnviRonment (SEER) \\	 \cline{2-4}
	
	&\cite{oyewumi2019isaac}& Recording network traffic/Poisoning Attack &Real Time Digital Simulator (RTDS) \\	 \cline{2-4}
	
	&\cite{kezunovic2019testbed}& Timing Intrusion Attack &Field End-to-End Calibrator/ Gold PMU \\ \cline{2-4}
	
	&\cite{marino2019cyber}& Test of cyber-physical sensor: IREST &Idaho CPS SCADA Cybersecurity (ISAAC) testbed \\ \cline{2-4}

	&\cite{konstantinou2019flep}& MITM attack/DoS attack & Open source software/Raspberry Pis. FLEP-SGS \\	
	
		  \hline

	\multirow{4}{*}{Cloud}& \cite{patil2019designing}&Flood malicious traffic (ICMP/HTTP/SYN)& VMware Esxi hypervisor/A vCenter server/VMs \\ \cline{2-4}
	
	& \cite{celesti2019approach}& Considering  small messages (about1–2 KBytes):  Fast filling of the buffers & MOM4Cloud architectural model. \\ \cline{2-4}
															
	&\cite{mishra2020kvminspector}& UNM database: Malicious tracing logs &KVM2.6.27 hypervisor/ Python3.4 \\ \cline{2-4}
	
    &\cite{van2016performance}& Test of memory usage before/after instance creation & OpenStack: Open-Source cloud operating system \\ \cline{2-4}
    
    &\cite{ullah2019design}&  Evaluation of performance metrics of NDN/edge cloud computing & Cloud VM  \\ \cline{2-4}
    
    &\cite{al2018remote}& Adding defaults: broken interconnection/Abnormal extruder &MTComm: Online Machine Tool Communication  \\ \cline{2-4}
    
    &\cite{sanatinia2017hyperdrive}& Side channel attacks/   stealthy data exfiltration &DHCP server/TFTP Server/HTTP Server/MQTT Server  \\ \cline{2-4}
    
    &\cite{frank2017design}& SQL Injection attack &OpenStack implementation/Python  \\ \cline{2-4}
    
    &\cite{gao2015cyber}& Testing  traffic scenarios & Openflow controller/OpenvSwitch/Network virtualization agent \\ \cline{2-4}
	
	&\cite{khorsandroo2018time}& Time inference attacks &Software Defined Network \\ \cline{2-4}	
	
	&\cite{kalliola2017testbed}& DDoS attack &OpenStack environment \\
		  \hline

\end{tabular}
\label{tab:Testbeds}
\end{center}
\end{table}

Ibrahim~\textit{et al.}~\cite{ibrahim2014modelling} proposed the use of a formal logic known as Secure Temporal Logic of Action [S-TLA.sup+] as a modelling-based approach for reconstructing evidence of Voice Over Internet Protocol~(VoIP) malicious attacks. The goal of the research was to generate related additional evidence and to measure the consistency against existing approaches using the [S-TLA.sup+] model checker.

Mace~\textit{et al.}~\cite{mace2018multi} reported on a multi-modelling-based approach to assessing the security of smart buildings. The approach was based on an Integrated Tool Chain for Model-based Design of Cyber-Physical Systems~(INTO-CPS), a suite of modelling, simulation, and analysis tools for designing cyber-physical systems. The study was motivated by the evolution to smart buildings controlled by multiple systems that provide critical services such as heating, ventilation, lighting, and access control, all highly susceptible to cyber-attacks. The stages of a systemic methodology to assessing the security when subjected to Man-in-the-Middle attacks on the data connections between system components by using a fan coil unit case study was presented.\\

\begin{table*}[hbt!]
\caption{\textbf{Types of Intrusion Detection Systems}}
\footnotesize
\centering
		\begin{tabular}{|p{0.5cm}|p{0.5cm}|p{3cm}|p{2.2cm}|p{3cm}|p{4cm}|}
		\hline
			 \textbf{Ref} & \textbf{Year}& \textbf{Specification}& \textbf{Domain}& \textbf{Software}& \textbf{Description}\\
			\hline
			\cite{al2020real} & 2020& Classical Signature detection& IoT& Cooja& Generation of DoS attacks\\
			\hline
			\cite{alazzam2020feature} & 2020& Classical Signature detection& -&Python& Pigeon Inspired Optimizer\\
			\hline
			\cite{kasongo2020deep} & 2020& RNN&Wirless technology& Python&IDS based on gated recruitment\\
			\hline
			\cite{mahdavi2020real} & 2020& Code-book of attack scenarios& -& C++/Linux environment& Using known attacks to detect anomaly\\
			\hline
			\cite{zhang2020model} & 2020&CNN& -& Python/LTS environment& Spatial-temporal feature detection\\
			\hline
			\cite{krzyszton2020simulation} & 2020& Machine learning&IoT& BMWatchSim&Anomaly detection by watchdogs\\
			\hline
			\cite{rajendran2019cross} & 2019& Rooting efficiency improvement&MANET&NS-2&Black hole attacks detection\\
			\hline
			\cite{abusitta2019deep} & 2019& Deep learning& Cloud&Python&IDS based on unsupervised training algorithm\\
			\hline
			\cite{suresh2020efficient} & 2019&Signature detection&WLAN &AC algorithm& Hardware approach architecture\\
			\hline
			\cite{kosmanos2020novel} & 2019&Machine learning&Electric vehicles&SUMO-OMNET-VEINS& Detection of spoofing attacks\\
			\hline
			\cite{zhang2019intrusion} & 2019&DRNN&IoT& MATLAB 2019b& Automated IDS for fog security\\
			\hline
			
				\hline
			\cite{selvakumar2019intelligent} & 2019&Deep learning&Vehicle security&Python& IDS for vehicle security by DNN\\
			\hline
			
		\cite{zhou2020distributed} & 2019&Signature detection&WSN&-& Selection based on Fuzzy logic\\
			\hline

		\cite{alyousef2019dynamically} & 2019&Signature detection&Vehicular Ad-Hoc&OMNET&Dynamic behavior analysis\\
			\hline

		\cite{condomines2019network} & 2019&Hybrid method&UAV Ad-hoc communication&JAVA& Spectral traffic analysis and robust controller\\
			\hline	
			
		\end{tabular}
	\label{tab:IDS types}
\end{table*}


\section{\textbf{The Future of CRs and TBs}}
\label{Section: TheFutureofCRsandTBs}

\subsection{\textbf{Future Trends}}
\begin{itemize}
    \item \textbf{Real-Time Auto-configurable Systems:}
    MIT's Lincoln Laboratory developed an advanced tool for cyber-ranges referred to as Automatic Live Instantiation of a Virtual Environment~(ALIVE),~\cite{braje2016advanced}, a range application extension to LARIAT. ALIVE has the capability of ingesting configuration files from Common Cyber Event Registration~(CCER) to automate the building out of Virtual Machines and networking infrastructure of the CR~\cite{9138883}. In addition to the capability to create virtual networks, it can also automate most of the system network build-outs, creating end hosts, routers, firewalls, and servers needed to support traffic generation. The host software packages and user accounts can also be installed. 
    
    The Cyber-Range Instantiation System~(CyRIS), an open source tool for facilitating cyber-range creation~\cite{pham2017automatic}, can execute efficient instantiation of cyber-ranges automatically. CyRIS automatically aids in the preparation and management of CRs using a pre-defined specification provided by the scenario managers or instructors. The tool contains both basic functions for establishing the infrastructure as well its security settings.\\
    
    ALPACA~\cite{eckroth2019alpaca} is one of the modern auto-configurable CR with the facility to set user-specified constraints to generate complex cyber-ranges. The core of the implementation are an AI planning engine, a database of vulnerabilities and machine specific configuration parameters with the ability to generate a VM that includes the sequences of vulnerabilities and exploits. \\
    
    \item\textbf{Smart, Mobile and Integrated Technologies:}
    Pharos~\cite{fok2011pharos}, a TB for Mobile Cyber-Physical Systems, is aimed at supporting mobile cyber-physical system evaluation in live networks. It is a networked system of independent mobile devices with its fundamental building block based on Proteus (an autonomous mobile system with highly modular software and hardware), with the capability of relating with each other and with networks of embedded sensors and actuators. Push-button repeatability facilitating the recreation of the same scenarios multiple times is an important feature of the TB.\\
    
    Cybertropolis~\cite{deckard2018cybertropolis} is aimed at breaking the paradigm of CRs and TBs by providing what is referred to as Cyber-electromagnetic~(CEMA) range facilities, which merges the features of CRs and TBs to yield a hybrid type of cyber-security training system. Cybertropolis was developed as a one-of-kind cyber-range that can be used in the areas of industrial control systems, cyber-physical devices, IoT and wireless systems. The platform provides the ability to create a heterogeneous network consisting of virtual Information and Communication Technology~(ICT) systems with integrated live cyber-physical systems, live Radio Frequency~(RF), and Internet of Things~(IoT) systems into a virtual environment.\\

    \item\textbf{Training with Augmented Reality Technology:} Augmented Reality (AR) is increasingly viewed as an important dimension in learning in different domains and is being considered as another impactful technology in future CR and TB training. AR offers the possibility of interaction with different parts of the systems, in so doing enriching the training owing to enhanced visualisation. Augmented reality TB or CR create a new interactive experience able to modify the trainee view of the progression of attacks. AR solution also gates portable solutions, as an example, the attack reaction could be modelled anywhere without infrastructure dependence. The environment can be modified and the programmer can add new attack scenarios.
\end{itemize}

\subsection{\textbf{Future Technologies}}
\begin{itemize}
    \item \textbf{5G/6G Technologies:}
   5th and 6th Generation (5G, 6G) networks will transform services using mobile and wireless network infrastructures by provisioning connections with advantageous features ranging from low latency with high network bandwidth capability through to machine-to-machine communication. 5G solutions enable better services using Virtualisation and Cloud technologies~\cite{tranoris2018enabling}, extending to Network Functions Virtualisation~(NFV) which enhances server virtualisation to network devices. Tranoris~\textit{et al.}~\cite{tranoris2018enabling}, utilised these capabilities to demonstrate real-time remote monitoring and video streaming between Vehicle-to-Vehicle~(V2V) in an assisted overtaking application~\cite{urquhart2019cyber}, showcasing the potential impact from emerging 5G and beyond. Mitra and Agrawal~\cite{mitra20155g}, described a highly futuristic connected society - ``smart living": Vehicle Ad-hoc Networks~(VANET) cloud for network connected transport systems managing dynamic real-time traffic demands; and  massive M2M communications. West~\cite{west20165g} also added that the revolution will bring about IoT-enabled health services while Letaief~\textit{et al.}~\cite{letaief2019roadmap} postulate that 6G will bring about ubiquitous AI-based services. The self-same capabilities present leveraging opportunities for CR and TB engineers and users to provide a seamless, faster, and low latency-based CR and TB deployments using virtual machines, sandboxes or containerised technologies.\\

    \item\textbf{Containerisation Technologies:}
    The impact of hypervisor-enabled virtualisation technology in CRs/TBs cyber-warfare training has been highly beneficial. VMs provide the required isolation from operational networks but present users with real-life training scenarios. The deployment of applications to implement VMs on data centres needs a dedicated guest operating system on each VM, on occasion different from the host operating system. Containerisation technology has been introduced as light-weight virtualised technology to that of VMs in order to manage these concomitant accrued overheads. A study conducted by Bhardwaj and Krishna~\cite{bhardwaj2019container} compared the use of the pre-copy VM migration scheme with that of the LXD/CR container migration technique, concluding that the use of latter reduces system downtime by 76.66\%, migration time by 65.55\%, scalability (volume of data transferred) by 76.63\%, throughput (number of transferred pages) by 76.78\%, overhead costs were reduced with regards to CPU utilisation by 55.89\% and RAM utilisation by 76.52\%. Thus, containerised technology costs less, guarantees more system up-time and saves times. Other studies that highlight the benefits of containerised technologies are Lovas~\textit{et al.}~\cite{lovas2019weather}, on their software container-based simulation platform in order to achieve scalability and portability; Mucci and Blumbers~\cite{mucci2019ted} to gain flexibility, reduce complexity while providing extensibility; and Kyriakou~\textit{et al.}~\cite{kyriakou2018container}, to ease deployment, management and resilience of their cloud-based environment. 
\end{itemize}

\subsection{\textbf{Future Application Areas}}
\begin{itemize}
    \item \textbf{Smart Cyber-Physical Systems:}
    Smart Cyber-Physical Systems~(sCPS) are large‐scale software intensive and pervasive systems, that are intelligent, self-aware, self-managing and self-configuring~\cite{delicato2020smart}. In line with other data driven artificially intelligence powered systems, sCPS utilise multiple data streams to manage real-world processes efficiently and through these offers a broad range of new applications and services in housing, hospital, transportation and automobile applications. \\ In recent times, cyber-criminals have up-skilled their skills through AI techniques to automate attacks, augment their strategies, launch more sophisticated attacks and by implication increase the success rates~\cite{kaloudi2020ai},~\cite{brundage2018malicious}. ICT tools and AI techniques have not only enriched the opportunities for cyber-criminals as a new form of threat landscape has suddenly emerged. There is a pressing obligation for cyber-range based training to evolve as a consequence implementing the detection as well as informing on optimum mitigation of these new threat dynamics. \\
    
    \item\textbf{Smart Cities and Industry 4.0:}
    The 4th Industrial revolution, also referred to as Industry 4.0, are data driven, network connected, digitalised industrial systems, heralding an era of automated manufacturing and service delivery with strong potential of process optimisation, imbued with new business practices. The evolution is, however, not without its attendant new cyber-threats. CyberFactory\#1~\cite{becue2018cyberfactory} is designed to proffer a solution between future digital factories and security threats gated by digitalisation. The principles on which the environment is established are conscious design, development, and demonstration of a System-of-Systems embracing the technical, economical, human and societal dimension of future factories~\cite{10.1145/3407023.3409313}. The platform demonstrates sets of major enabling capabilities that foster optimisation and resilience of next generation manufacturing and service delivery industries. As the evolution unfolds,  there is a need to continue to propose new solutions capable of mitigating the dilemma between the deployment of future factories/smart cities and cyber-threats.
    
    A body of available literature stresses that cyber threats and privacy concerns will increase significantly in smart systems due to high degrees of network inter-connectivity; Reys~\textit{et al.}~in~\cite{reys2016smart}, Baig~\textit{et al.}~in~\cite{baig2017future}, Vitunskaite~\textit{et al.}~in~\cite{vitunskaite2019smart}, Mylrea~\textit{et al.} in~\cite{mylrea2015singapore}, Srivastava~\textit{et al.} in~\cite{srivastava2017safety}, Aldairi~\textit{et al.} in~\cite{aldairi2017cyber}, Cerrudo~\textit{et al.} in~\cite{cerrudo2015hacking}, Alibasic~\textit{et al.} in~\cite{alibasic2016cybersecurity} and Braun~\textit{et al.} in~\cite{braun2018security}. Wang~\textit{et al.}~in~\cite{wang2015data} and Farahat~\textit{et al.}~in~\cite{farahat2019data} focused on data security as well as threat modeling for smart city infrastructures. Vattapparamban~\textit{et al.} in~\cite{vattapparamban2016drones} expect that drones will be used in service delivery in highly connected smart cities environments of the future and hence will become a factor in defining the scope of cyber-attacks. Li~\textit{et al.}~in~\cite{li2017deployment} report on the intelligent management of network traffic to avoid congestion while reducing cyber-security concerns in Smart cities. \\

    \item\textbf{Aerospace and Satellite Industries:} The evolution of the aerospace and satellite industries and the significant contribution the sector makes to the health of the economy has made them a central interest for cyber-attacks. CRs and TBs are essential to model the impact of cyber-attack effects and enhance the ability of protecting this critical infrastructure. The goal is to understand and overcome the spectrum of possible attacks by taking into account the sensitivity of information used. Virtualisation using a simulation-based system is a potential solution to implement TBs, but a total recognition of several parts of such critical infrastructure should be studied. The prediction of the hacker's strategies and aims remain the core to understanding the optimum countermeasure against class of attack.\\
\end{itemize}

\section{\textbf{Conclusion}}
\label{Section: Conclusions}
The rapid proliferation in the automation of cyber-attacks is diminishing the gap between information and operational technologies and in turn stimulating an increased reliance on training to inculcate robust cyber-hygiene knowledge for cyber-security professionals, trainers and researchers. Cyber-Situational awareness is now viewed as a central spine in the effective provision of practices that protect organisations/infrastructures against a cohort of more sophisticated cyber-attackers. From necessity, the training must be delivered through non-operational environments that provide real-time information on cyber-threats, their early identification/characterisation and effective countermeasures. This paper presents an evaluation of prominent CR and TB platforms segmented by type, technology, threat scenarios, applications and the scope of attainable training. Furthermore, a novel taxonomy for CRs and TBs is presented which represents the foundation for the prediction of the evolution of CRs/TBs. In all, this automation has accentuated a rapidly diminishing differentiation between CRs and TBs respective areas of application.

 \vspace{6pt} 




\funding{{\huge\euflag} The research is supported by the European Union Horizon 2020 Programme under Grant Agreement no. 833673. The content reflects the authors’ view only and the Agency is not responsible for any use that may be made of the information within the paper.
}

\conflictsofinterest{The authors declare no conflict of interest}

\reftitle{References}
\bibstyle{spbasic.bst}
\bibliography{biblio_traps_dynamics}

\end{document}